\documentclass[preprint]{elsarticle}

\usepackage{lineno,hyperref}

\usepackage{amsmath,amssymb}
\usepackage[ruled,vlined]{algorithm2e}
\usepackage{graphicx}

\usepackage{subcaption}

\usepackage{sistyle} % for degree symbol \arcdeg and micro symbol

\newcommand{\argmin}{\operatornamewithlimits{argmin\ }}
\newcommand{\argmax}{\operatornamewithlimits{argmax\ }}

\newcommand{\real}{\Re}

\newcommand{\numPixels}{N}
\newcommand{\numTimePointsBeforCompression}{L}
\newcommand{\numTimePointsAfterCompression}{k}

\def\dictCompValue{r}
\def\autocalliCompValue{\numTimePointsAfterCompression}

\newcommand{\numSamplesPerRO}{S}

\newcommand{\regterm}{\mathcal{R}}
\newcommand{\samplOp}{\mathcal{G}}
\newcommand{\dictManifold}{\mathcal{D}}
\newcommand{\compOp}{\mathcal{C}}
\newcommand{\dictProj}{\mathcal{P}_{ \dictManifold_c }}

\newcommand{\samplPattern}{\mathcal{M}}
\newcommand{\fourierOp}{\mathcal{F}}
\newcommand{\coilSensOp}{\mathcal{S}}

\newcommand{\affect}{:=}
\newcommand{\iter}{k}

\def\leftMat{\widetilde{\mathbf{U}}}
\def\rightMat{\widetilde{\mathbf{V}}^H}

\def\righMatT{\mathbf{V}}

\def\compMat{\widetilde \righMatT}

\def\dicCompMat{\widetilde{\righMatT}_d}
\def\autocalCompMat{\widetilde\righMatT_{ac}}

\def\autocalCompU{\mathbf{U}_{ac}}
\def\autocalCompS{\mathbf{S}_{ac}}
\def\autocalCompV{\mathbf{V}_{ac}}

\def\dicMat{\mathbf{D}}

\DeclareMathOperator{\prox}{prox}
\DeclareMathOperator{\rank}{rank}

\def\methodXXIV{IGP-MRF-01}
\def\methodXXXIV{IGP-MRF-00}
\def\methodGFB{GFB-MRF}
\def\methodXXXIX{IGP-MRF-11}
\def\methodXXXVIII{IGP-MRF-10}
\def\ZhaoLR{Zhao-LR}

\modulolinenumbers[5]

\journal{Journal of \LaTeX\ Templates}

\begin{document}

\begin{frontmatter}

\title{\methodGFB: A parallel spatial and Bloch manifold regularized iterative reconstruction method for MR Fingerprinting}
\tnotetext[mytitlenote]{The concepts and information presented in this paper are based on research results that are not commercially available.}

\author[1]{Simon Arberet \corref{cor1}} % \fnref{fn1}
\ead{simon.arberet@siemens-healthineers.com}

\author[1]{Xiao Chen}
\author[1]{Boris Mailh{\'e}}
\author[2]{Peter Speier}
\author[2]{Gregor K{\"o}rzd{\"o}rfer}
\author[2]{Mathias Nittka}
\author[2]{Heiko Meyer}
\author[1]{Mariappan S. Nadar}



\address[1]{Digital Technology \& Innovation, Siemens Healthineers, Princeton, NJ, USA}
\address[2]{Magnetic Resonance, Siemens Healthineers, Erlangen, Germany}

\begin{abstract}
Magnetic Resonance Fingerprinting (MRF) reconstructs tissue maps based on a sequence of very highly undersampled images.
In order to be able to perform MRF reconstruction, state-of-the-art MRF methods rely on priors such as the MR physics (Bloch equations) and might also use some additional low-rank or spatial regularization. 
However to our knowledge these three regularizations are not applied together in a joint reconstruction. 
The reason is that it is indeed challenging to incorporate effectively multiple regularizations in a single MRF optimization algorithm.
As a result most of these methods are not robust to noise especially when the sequence length is short.
 %to regularize the images in the temporal dimension, but usually do not exploit in addition the spatial structure of the images. 
%Most of MRF algorithms that exploit more than one type of regularization, are using either spatial and Bloch manifold regularizations, low-rank and Bloch manifold regularizations, or spatial and low-rank regularizations but not the three together.
% spatial regularization effectively in addition to the Bloch manifold and/or low-rank regularization(s) in an MRF algorithm.
In this paper, we propose a family of new methods where spatial and low-rank regularizations, in addition to the Bloch manifold regularization, are applied on the images. 
We show on digital phantom and NIST phantom scans, as well as volunteer scans that the proposed methods bring significant improvement in the quality of the estimated tissue maps.
%The spatial regularization and the Bloch manifold regularization are applied on the images, in parallel, and then merged by an averaging. This can be interpreted as an application of the generalized forwad-backward algorithm (GFB).
\end{abstract}

\begin{keyword}
Magnetic resonance imaging \sep Magnetic Resonance Fingerprinting \sep image reconstruction \sep iterative reconstruction
\end{keyword}

\end{frontmatter}

%\linenumbers

\section{Introduction}

% What is the problem we are trying to solve ?
Magnetic  Resonance (MR) Fingerprinting (MRF) is a recent technique \cite{ma2013magnetic} for estimating quantitative tissue maps which uses a non-stationary pulse sequence
(i.e. the sequence of parameters such as repetition time (TR), flip angle (FA), and inversion time, is changing in each time point) 
to generate a sequence of continuously changing image contrasts, acquired at a high under-sampling ratio. 
%and heavily undesampled MR acquisition. 
% (i.e. the sequence of parameters such as repetition time, flip angle, and inversion time, is changing in each time point) resulting in a different contrast evolution
The name \textit{fingerprinting} comes from the fact that, in order to recover the quantitative parameters, the acquired contrast signals are matched against a dictionary of precomputed magnetization signal responses called \textit{fingerprints}, by analogy with fingerprints used for forensics.
These fingerprints, obtained via Bloch equation simulations, model the possible signal evolutions, given the MR parameter sequence and the possible combinations of tissue parameters (T1, T2, proton density, etc.) expected to be present in the scanned object.
The MR sequence uses a non stationary sequence of parameters (FA, TR, sampling pattern, etc.) in order to create spatial and temporal incoherence among the signal responses from different tissue types.
While the sequence used in the original MRF article \cite{ma2013magnetic} was random or pseudo-random, later results \cite{zhao2016optimal} where the sequence was optimized, showed a resulting highly structured sequence. 
% why is it interesting ? (why you should care about it ?)
%MRF has the advantage of providing quantitative images of the important tissue properties, as opposed to the classical approach of MR imaging which only provides weighted images of the magnetization response which weighting depends on many factors such as the scanner, the detector used, etc. 
%The quantitative images obtained via MRF provide useful information for 

% What is the problem we are trying to solve ?
MRF provides an effective solution for quantitative imaging, but requires a highly undersampled acquisition in order to have a reasonable scanning time. 
The classical MRF technique \cite{ma2013magnetic} obtains the tissue maps by first zero-filling the undersampled k-space and then by performing the fingerprint matching.

Following the initial MRF method \cite{ma2013magnetic}, new techniques \cite{mcgivney2014svd,cao2017robust} were proposed which exploit the redundancy of the fingerprints in the time domain and combining different time points prior to the fingerprint matching in order to ``dealiase'' the images before the fingerprint matching step.
%(2)
% what are the other solutions and why they aren't satisfactory ?
Then, some approaches inspired by compressed sensing \cite{davies2014compressed,cline2017air,wang2016magnetic, zhao2016maximum} improved further the performance of MRF by iteratively removing artefacts by alternating between Bloch manifold projection (aka dictionary matching) and gradient update. In order to compensate for numerical instabilities in the high frequency corners of the k-space where the sampling pattern (usually spiral or radial) is not taking measurements, which can cause checkboarding artefacts, these methods may use an additional low-pass filter at each iteration \cite{pierre2016multiscale, cline2017air}. Note that the method described in \cite{pierre2016multiscale} alternates between dictionary matching and low-pass filtering, where the low-pass filter changes accross iterations in order to gradually include the high frequencies.
However, these methods did not (or not successfully \cite{davies2014compressed,wang2016magnetic}) exploit spatial regularities which are naturally present in images.

On the other hand, low-rank priors have been exploited \cite{doneva2017matrix, zhao2017improved, mazor2016low, asslander2018low, bustin2019high,lima2019sparsity} in order to capture redundancy of the MRF signal in the temporal dimension. 
These methods are iterative and often computationaly intensive as they require to compute multiple singular value decompositions (SVD) \cite{ mazor2016low,asslander2018low} or higher-order SVD \cite{bustin2019high}.
Note that the recent patch-based method HD-PROST \cite{bustin2019high} exploits correlated structure of the spatio-temporal image through local (within a patch), non-local (between similar patches) and contrast (i.e. time) scale using higher-order singular value decomposition (HOSVD). In a sense, this last method exploits some sort of spatial structure but the Bloch manifold (dictionary matching) is not used as a prior in the optimization problem, but only at the end-step in order to generate the T1 and T2 maps from the reconstructed contrast images.
The recent method \cite{lima2019sparsity} is iteratively and jointly optimizing (locally) low-rank and spatial priors, but as in \cite{bustin2019high}, the dictionary matching is not used as a prior in the optimization problem but only as a final step to convert the MRF image to the  tissue parameters.
%but these solutions are often computationaly intensive, limited to low resolution images and not so effective. % as we will see in the experiments section (section \ref{sec_experiments}).
%However these methods do not exploit spatial regularities which are naturally present in images as well as low-rank regularization.
%Some other methods proposed to exploit low-rank regularization [Zhao, FLOR, Doneva], or spatial regularization [Wang, Davies tried], but as we will see, these approaches are often very inefficients [FLOR, Doneva] and do not improve the state-of-the art [BLIP, AIR-MRF].

%(3)
% what is our solution ?
% compare to the other solutions and why is it better ?

In this paper, which extends results presented in \cite{arberet2017} and \cite{arberetGFB2019}, we propose an efficient iterative reconstruction method which, in addition to the Bloch manifold regularization, incorporates low-rank and spatial regularizations.
% and which improves significantly the quality of the reconstruction compared to the state of the art.
We propose two optimization algorithms to combine these three regularizations (Bloch manifold, low-rank and spatial) in a single algorithm: an incremental gradient proximal algorithm \cite{bertsekas2011incremental} where the regularizations are applied sequentially, and a generalized forward-backward algorithm \cite{raguet2013generalized} where the regularizers are applied in parallel.
The low-rank prior is not applied as a nuclear norm minimization which would be computationally expensive, but via dictionary SVD compression \cite{mcgivney2014svd,cline2017air} which is computed offline. 
As in \cite{zhao2017improved}, we set a threshold much below the rank of the dictionary which provides low-rank regularization (at the expense of a negligible increase in the approximation error) and greatly improves computational efficiency of the algorithm. 
As we will discussed in \ref{sec_lowrank} (see also the discussion in \cite{zhao2017improved}), decreasing the rank beyond the dictionary rank has the effect of improving low-rank regularization as the expense of linear projection approximation.
%A simple and efficient way to impose a low-rank prior, as opposed to nuclear norm minimizarion,  is to simply compress the dictionary via SVD as in \cite{mcgivney2014svd,cline2017air} but at a much stronger rate, much below the rank of the dictionary (as in \cite{zhao2017improved}).
Another approach studied in this paper is to compute a low-rank subspace via auto-calibration. In both cases, the approaches have the consequence of reducing the temporal dimension of the optimization variables. 
Indeed, thanks to this temporal compression, the spatial regularization can now be performed very efficiently in the compressed domain.
In this paper we use 2D total variation (TV) as an example regularization via a few iterations of the Chambolle-dual algorithm \cite{chambolle2004algorithm}, but other spatial regularizers such as cycle spinning \cite{kamilov2014variational,mailhe2018fast} or joint 
(along the compressed-temporal dimension) could also be applied. We actually tried these last two regularizers in another (internal) study and obtained similar results, however it is beyond the scope of this paper to evaluate different spatial regularizers. 
% and provides similar quality results. Joint (along the compressed-temporal dimension) sparse regularization also provides good performance. 
The fingerprints matching, also performed in the compressed domain, can be further accelerated via e.g. a kd-tree algorithm as in \cite{cline2017air}, FGM \cite{cauley2015fast}, or other approximate nearest neighbors (ANN) techniques \cite{muja2009flann} or deep neural networks (DNN) \cite{cohen2017deep,fang2019deep,song2019magnetic}, and/or even be skipped for inner iterations as we will see later in section \ref{sec_incremental}.

\paragraph{Contributions}
\begin{itemize}
\item Our main contribution is to propose two new algorithms for MRF reconstruction which exploit jointly and iteratively three regularizations: Bloch manifold regularization, spatial regularization, and low-rank.
To our knowledge it is the first time that these three regularizations are optimized jointly. In particular in \cite{davies2014compressed,  cline2017air, zhao2016maximum, pierre2016multiscale,mazor2016low,asslander2018low}, no spatial regularization is performed, and in \cite{wang2016magnetic, doneva2017matrix,  mazor2016low, bustin2019high, lima2019sparsity} the Bloch manifold projection (i.e. dictionary matching) is not performed iteratively but only as a final stage to convert the MRF image to tissue maps.
\item We also propose an autocalibration mechanism to impose low-rank in our iterative algorithm so that the low-rank subspace to be data-driven rather than dictionary driven as in \cite{mcgivney2014svd}. 
\item We compare the proposed variants of our algorithms on a series of experiments: evaluation on synthetic Brainweb data w.r.t. the sequence length (from 600 TR to 3000 TR) in the cases with and without added noise to the measurements in order to test the robustness of these algorithms.
\item After selecting our best candidate among our proposed methods, we compare our (best) method against four different methods from the state-of-the-art on a series of experiments: on synthetic Brainweb data w.r.t. the sequence length (from 600 TR to 3000 TR) with and without added noise to the measurements, and also on a real phantom scan and a real volunteer scan, and in both cases with and without added noise.
\end{itemize}
     
The outline of this paper is as follows: We first present our MRF methods in section \ref{sec_method}, i.e. the optimization problem we are trying to solve and our proposed algorithms, the implementation of the low-rank prior, the fingerprint matching, the spatial regularization, and the step-size selection. Then in section \ref{sec_experiments}, we present results of experiments, first on Brainweb digital phantom data for which we have a ground truth and thus can provide quantitative results, and then on NIST phantom \cite{nistphantom} and volunteer data. 
We finish by a discussion on the results and a conclusion.

%Recently some approaches using a compressed sensing approach improved the performance of MRF. However 

%(1)
% What is the problem we are trying to solver ?
% why is it interesting ? (why you should care about it ?)

%(2)
% what are the other solutions and why they aren't satisfactory ?

%(3)
% what is our solution ?
% compare to the other solutions and why is it better ?

%(4)
% discuss related work where similar techniques and experiments have been used and applied to different problems
% motivation for low-rank and TV...

% what are the contributions ?
% what is really new (and what isn't) ?

% what is the context of our problem ?

% what is our solution ?

% how our paper differs and improves upun previous work ?

\section{Method\label{sec_method}}
%The proposed methods include 1)  a low-rank regularization via dictionary compression and 2) a spatial regularization via TV minimization of each compressed image. 

The problem we want to solve is to reconstruct the MRF image $\mathbf{X} \in \mathbb{C}^{\numPixels \times \numTimePointsBeforCompression}$ assuming a discretization into $\numPixels$ voxels, from the observed sequence of length $\numTimePointsBeforCompression$ of k-space samples $\mathbf{Y} \in \mathbb{C}^{\numSamplesPerRO \times \numTimePointsBeforCompression}$, where $\numSamplesPerRO$ is the number of samples per readout. In a second step, the quantitative parameters (T1, T2, ...) can then be retrieved from this image by fingerprint matching (or other regression technique) as described in section \ref{sec_fingerMatch}.

As in AIR-MRF \cite{cline2017air}, we rely on a SVD compression dictionary technique \cite{mcgivney2014svd} which exploits the compressible (low-rank) structure of the dictionary and has the additional advantages of 1) reducing the size of the optimization variables and thus accelerates and simplifies the optimization problem, 2) accelerating the Fourier transform and gridding operations which are the most costly operations of the algorithm when the k-space trajectory is non Cartesian and multiple coils are used. Indeed, as compression commutes with the Fourier transform and gridding operations, the two operations can be performed much more efficiently in the compressed domain of size $\numTimePointsAfterCompression$  ($\numTimePointsAfterCompression \ll \numTimePointsBeforCompression$) than in the original domain of length $\numTimePointsBeforCompression$.
% as compression commutes with the Fourier and gridding operations, which are some of the most costly operations of the algorithm when the k-space trajectory is non cartesian and multiple coils are used, this last two operations can be performed in the compressed domain and thus much more efficiently.  
As we will see later, SVD compression has other advantages. % Rq it also allow a better conditioning of the NUFFT because the data is fully sampled after compression, kd-tree works also better on PCA data, accelerating the matching

Let's denote the MRF images after compression by $\mathbf{X}_c  \triangleq \compOp (\mathbf{X}) \in \mathbb{C}^{\numPixels \times \numTimePointsAfterCompression}$.
Our approach consists of first solving optimization problem \eqref{eq_optimProblem} to estimate the denoised MRF images, and then retrieving the corresponding tissue parameters via fingerprint matching.
Optimization problem \eqref{eq_optimProblem} is:
\begin{align}\label{eq_optimProblem}
\argmin_{\hat{\mathbf{X}}_c \in \mathbb{C}^{\numPixels \times \numTimePointsAfterCompression}} \ & \| \mathbf{Y} -  \samplOp ( \hat{\mathbf{X}}_c ) \|_2^2 + \lambda \regterm(\hat{\mathbf{X}}_c) \\
\text{subject to}\ &\hat{\mathbf{X}}_{c (i,:)} \in \dictManifold_c,\  i=1,\ldots,\numPixels, \nonumber
%\text{minimize}_{\hat{X}} \|| Y - G( \hat{X} ) \||_2^2 + \lambda R(\hat{X})
\end{align}
where $\samplOp \triangleq \samplPattern \fourierOp \coilSensOp \compOp^H = \samplPattern \compOp^H \fourierOp \coilSensOp$ is the observation operator, $\coilSensOp$ the coil sensitivity map, $\fourierOp$ the Fourier transform, and $\samplPattern$ the undersampling operator. 
Note that, as shown in \cite{asslander2018low}, the compression operator $\compOp$ commutes with the Fourier operator, and thus the number of Fourier transforms can be reduced drastically by performing them in the compressed domain. %$\samplOp = \samplPattern \compOp^H \fourierOp \coilSensOp$
In our implementation, sampling is performed with a series of single-shot spiral trajectories in k-space, and $\samplPattern$ and $\fourierOp$ together are implemented with a non-uniform fast Fourier transform (NUFFT) \cite{fessler2003nonuniform}.

$\regterm$ is a regularization operator which in our case enforces spatial regularization e.g. using a total variation (TV) regularization. 
$\dictManifold_c$ is the cone of the Bloch response manifold after compression. The Bloch manifold is indeed a cone as it is stable to nonnegative scaling (the proton density in our case) \cite{davies2014compressed}, and its SVD compression is a linear projection of a cone which is still a cone. 
 %  i.e. the set of signal $\mathbb{R}_+ \mathcal{B}$ , where $\mathcal{B}$ is the
%$\dictProj$ is the projection onto the dictionary of Bloch signal responses.
This Bloch manifold projection will be denoted by $\dictProj$ in the rest of the paper and the details of this projection will be discussed in section \ref{sec_fingerMatch}.
Note that the Bloch manifold is not convex, and moreover its projection is usually implemented via fingerprint matching, i.e. a non-smooth (non differentiable) projection. 

In the following, we propose different variants of the CS-MRF method. These variants and their names are summarized in Table \ref{tab_proposedmthods}.

\subsection{Incremental gradient proximal MRF (IGP-MRF)\label{sec_incremental}}

In order to solve optimization problem \eqref{eq_optimProblem}, we can rely on the \textit{incremental (sub)gradient-proximal method} (IGP) %[Bertsekas eq (4.14 and 4.15) of Optimization for Machine Learning, or 
\cite[equations 2.3 and 2.4]{bertsekas2011incremental}. % which applied to our problem leads to algorithm \ref{algo_IncGradProx}.
% The method is called Incremental subgradient-proximal method, but here: 1) there is only one term i, and 2) the function is differentiable.
Algorithm \ref{algo_IncGradProx} is an IGP algorithm that optimizes the image volume in the compressed domain,  as well as the tissue parameters $\hat{\theta} = \{ \text{T}1, \text{T}2, \ldots \}$ and the proton density $\hat{\rho}$, which are obtained during the fingerprint matching step (will be described in section \ref{sec_fingerMatch}).
The strategy for the gradient step size $\alpha_{\iter}$ will be discussed in section \ref{sec:stepSizeSel}. Note that $\samplOp^H$ denotes the adjoint of operator $\samplOp$.
As will be explained in section \ref{sec_lowrank}, the low-rank prior imposes the solution to be in a low-dimensional subspace which approximates the Bloch manifold.
As a consequence, this prior might be enough to replace the fingerprint matching as a regularizer (in a similar way as in \cite{mazor2016low} and \cite{zhao2017improved}). In that case we could skip the fingerprint matching except at the final iteration to retrieve the tissue maps.
Note that if the fingerprint matching/ Bloch manifold projection is skipped in optimization problem \eqref{eq_optimProblem}, then this problem becomes smooth and convex (as only the Bloch manifold projection is non-convex and non-smooth) for which many algorithms and nice convergence properties exist. 
Then Algorithm \ref{algo_IncGradProx} with skipped fingerprint matching reduces to the well  known (convex) gradient-proximal algorithm.
In practice we also notice that, in the case we drop the fingerprint matching in the iterative procedure, it is still useful to have the first matching at the first iteration as it provides a good initialization.
The variants without matching are denoted \methodXXXIV\ and \methodXXXVIII\ in Table \ref{tab_proposedmthods}.
These methods were first  introduced in our ISMRM abstract \cite{arberet2017} where they were denoted CS-MRF.

 \begin{algorithm}  [t!]          % enter the algorithm environment
\SetAlgoLined \KwIn{$\mathbf{Y}$, $\lambda$.} 
\text{Initialize:} \\
$\iter \affect 0$, $\hat{\mathbf{X}}_c^{(0)} \affect 0$.\\
\While{$\iter < K_{max}$ }{
\textbf{Gradient update:} \\
$\mathbf{Z}_c \affect \hat{\mathbf{X}}_c^{(\iter)} -\alpha_{\iter} \samplOp^H(\samplOp(\hat{\mathbf{X}}_c^{(\iter)}) -  \mathbf{Y} ) $\\
\textbf{Fingerprint matching*:} \\
$[\mathbf{Z}_c, \hat{\theta}, \hat{\rho}] \affect \dictProj(\mathbf{Z}_c )  $\\
\textbf{Spatial regularization:} \\
$\hat{\mathbf{X}}_c^{(\iter+1)} \affect \prox_{\alpha_{\iter} \lambda  \regterm} (\mathbf{Z}_c)  $\\
$\iter \affect \iter + 1$\\
}
\Return $\hat{\mathbf{X}}_c^{(\iter)},  \hat{\theta}, \hat{\rho}$ \\
\caption{IGP-MRF algorithm \label{algo_IncGradProx} \newline * for methods \methodXXIV\ and \methodXXXVIII, the fingerprint matching step is applied only in the first and last iteration.} 
\end{algorithm}

% \begin{algorithm}  [t!]          % enter the algorithm environment
%\SetAlgoLined \KwIn{$\mathbf{Y}$, $\lambda$.} 
%\text{Initialize:} \\
%$\iter \affect 0$, $\hat{\mathbf{X}}_c^{(0)} \affect 0$.\\
%\While{$\iter < K_{max}$ }{
%\textbf{Gradient update:} \\
%$\hat{\mathbf{X}}_c^{(\iter+1/3)} \affect \hat{\mathbf{X}}_c^{(\iter)} -\alpha_{\iter} \samplOp^H(\samplOp(\hat{\mathbf{X}}_c^{(\iter)}) -  \mathbf{Y} ) $\\
%\textbf{Fingerprint matching*:} \\
%$[\hat{\mathbf{X}}_c^{(\iter+2/3)}, \hat{\theta}, \hat{\rho}] \affect \dictProj(\hat{\mathbf{X}}_c^{(\iter+1/3)} )  $\\
%\textbf{Spatial regularization:} \\
%$\hat{\mathbf{X}}_c^{(\iter+1)} \affect \prox_{\alpha_{\iter} \lambda  \regterm} (\hat{\mathbf{X}}_c^{(\iter+2/3)})  $\\
%$\iter \affect \iter + 1$\\
%}
%\Return $\hat{\mathbf{X}}_c^{(\iter)},  \hat{\theta}, \hat{\rho}$ \\
%\caption{Incremental CS-MRF algorithm \label{algo_IncGradProx} \newline * for methods \methodXXIV\ and \methodXXXVIII, the fingerprint matching step is applied only in the first and last iteration.} 
%\end{algorithm}

% step size
% stopping criteria

\subsection{Generalized forward-backward MRF (\methodGFB)}

We propose another algorithm where the fingerprint matching and the spatial regularization are performed in parallel rather than sequencially.
The algorithm is derived from the generalized forward-backward splitting algorithm (GFB) \cite{raguet2013generalized} which generalizes the forward-backward algorithm to deal with multiple non-smooth functions (which are two in our case: the indicator function of the Bloch manifold, and the TV norm). As for the forward-backward algorithm, the regularity of the data fidelity term is explicitly used with a gradient step (forward step), and the proximity operators of the non-smooth terms are applied in parallel (backward step).
GFB algorithm has nice theoretical properties \cite{raguet2013generalized} such as a proven convergence with robustness to errors on the computation of the proximity operators and errors of the gradient of the smooth term. Convergence of GFB are also proven for constant or decreasing step-size. Note that in our case the assumptions of the GFB algorithm are not met, in particular the Bloch manifold is not convex.
This method was first introduced in our ISMRM abstract \cite{arberetGFB2019}.

 \begin{algorithm}  [t!]                   % enter the algorithm environment
\SetAlgoLined \KwIn{$\mathbf{Y}$, $\lambda$.} 
\text{Initialize:} \\
$\iter \affect 0$, $\hat{\mathbf{X}}_c^{(0)} \affect 0$, $\mathbf{Z}_{bloch}^{(0)} \affect 0$, $\mathbf{Z}_{spat}^{(0)} \affect 0$.\\
\While{$\iter < K_{max}$ }{
\textbf{Gradient update:} \\
$\mathbf{G} \affect \hat{\mathbf{X}}_c^{(\iter)} -\alpha_{\iter} \samplOp^H(\samplOp(\hat{\mathbf{X}}_c^{(\iter)}) -  \mathbf{Y} ) $\\
% \text{Offset update:} \\
% $\mathbf{O}_{bloch} \affect \hat{\mathbf{X}}_c^{(\iter)} -  \hat{\mathbf{Z}}_{bloch}^{(\iter)} $, $\mathbf{O}_{spat} \affect \hat{\mathbf{X}}_c^{(\iter)} -  \hat{\mathbf{Z}}_{spat}^{(\iter)} $\\
\textbf{Fingerprint matching:} \\
$\mathbf{Z}_{bloch}^{(\iter+1)} \affect \dictProj( \mathbf{G} +   \mathbf{O}_{bloch})  -  \mathbf{O}_{bloch}$\\
\text{with}  $\mathbf{O}_{bloch} = \hat{\mathbf{X}}_c^{(\iter)} -  \mathbf{Z}_{bloch}^{(\iter)} $\\
\textbf{Spatial regularization:} \\
$\mathbf{Z}_{spat}^{(\iter+1)} \affect \prox_{\alpha_{\iter} \lambda  \regterm}( \mathbf{G} +   \mathbf{O}_{spat})  -  \mathbf{O}_{spat}$\\
\text{with} $\mathbf{O}_{spat} = \hat{\mathbf{X}}_c^{(\iter)} -  \mathbf{Z}_{spat}^{(\iter)} $\\
\textbf{Merging:} \\
$\hat{\mathbf{X}}_c^{(\iter+1)} \affect \left( \mathbf{Z}_{bloch}^{(\iter+1)} +  \mathbf{Z}_{spat}^{(\iter+1)}  \right) \Big/2$\\
$\iter \affect \iter + 1$\\
}
\textbf{Fingerprint matching:} \\
$[\hat{\mathbf{X}}_c^{(\iter+1)}, \hat{\theta}, \hat{\rho}] \affect \dictProj(\hat{\mathbf{X}}_c^{(\iter)} )  $\\
\Return $\hat{\mathbf{X}}_c^{(\iter+1)}, \hat{\theta}, \hat{\rho}$ \\
\caption{ \methodGFB\ algorithm \label{algo_GFB}} 
\end{algorithm}

\subsection{Low-rank prior\label{sec_lowrank}}
There exists different ways to impose a rank constraint on the data in a matrix form (sometime called Casorati matrix) $\mathbf{X} \in \mathbb{C}^{\numPixels \times \numTimePointsBeforCompression}$.
One way is to project the data on the non-convex set of low-rank matrices (i.e. $\{ \mathbf{X}\ | \rank(\mathbf{X}) \leq  \autocalliCompValue \}$), using the Eckart-Young theorem and performing a SVD decomposition of the matrix followed by an singular value hard thresholding. Another way is to relax this non-convex constraint to its convex hull, i.e. replacing the rank constraint by the nuclear norm, defined as the sum of the singular values of the matrix. While this approach is convex, it still relies on a costly SVD followed by a soft-thresholding operation. This approach was used in \cite{mazor2016low}.

\paragraph{Low-rank via SVD compression \label{sec_lowranksvd}}
Another way to impose low-rank is via a low-rank factorization of $\mathbf{X}$ \cite{recht2010guaranteed, zhao2017improved}, i.e. instead of searching for an optimal matrix $\mathbf{X}$ having a low-rank,  we are searching for two matrices $\leftMat \in \mathbb{C}^{\numPixels \times \autocalliCompValue}$ and $\rightMat \in \mathbb{C}^{\autocalliCompValue \times \numTimePointsBeforCompression}$ with $\autocalliCompValue < \numPixels$ and $\autocalliCompValue < \numTimePointsBeforCompression$, and such that  $\mathbf{X} = \leftMat \rightMat$.
%The main advantage of this approach compared to the first one, is that it decreases significantly the size of the optimization variables. 
While optimizing jointly the two matrices $\leftMat$ and $\rightMat$ is not convex, in some problems one of these two matrices can be fixed with some a priori knowledge. In that case the problem is simplified and becomes convex.
%, decreasing its number of variables and imposing the low-rank constraint in a simple and convex way.
%In particular, in the MRF context, the rows of $\leftMat$ span the spatial subspace of $\mathbf{X}$ while the columns of $\rightMat$ span the temporal subspace of $\mathbf{X}$.
As the rows of $\mathbf{X}$ are supposed to be in the cone of the Bloch response manifold according to the MR physics, and that this cone is contained in a relatively low dimensional subspace, the principal temporal components of the Bloch signal dictionary $\dicCompMat$ can be used to form the matrix $\compMat$ (conjugate transpose of $\rightMat$). 
In this case, $\mathbf{X}_c \triangleq \compOp (\mathbf{X}) = \compOp_d (\mathbf{X})  \triangleq \mathbf{X}  \dicCompMat$, where $\dicCompMat$ is obtained by keeping the $\autocalliCompValue$ first right singular vectors (i.e. the $\autocalliCompValue$ first columns of $\mathbf{V}_d$) of the dictionary $\dicMat$, where $\dicMat = \mathbf{U}_d \mathbf{S}_d {\mathbf{V}_d}^H$ is the SVD decomposition of $\dicMat$.
This idea was used in previous work \cite{cline2017air}, where SVD compression was used with a rather large rank (200, i.e. larger than the rank of the dictionary) in order to reduce dimensionality and speeding up the NUFFT and matching time, but not to impose a low-rank constraint.
In \cite{zhao2017improved} this idea of imposing a low-rank constraint via this approach was used with a rank value of around $\autocalliCompValue = 6$ to $10$. In this paper we used a value of $\autocalliCompValue = 10$ for all our proposed methods (shown in Table \ref{tab_proposedmthods}) and experiments.
To choose this value, we performed experiments for different values of $\autocalliCompValue$ and noticed that the best performance were in the range 10 to 20, and started to degrade for values smaller than 10. So as smaller the value faster the reconstruction, the value of $\autocalliCompValue=10$ was a selected.
%critics ...

All of our proposed methods in Table \ref{tab_proposedmthods} use a dictionary SVD compression.
We also propose in the next paragraph another approach which in addition to the dictionary SVD compression exploits auto-calibration data to impose a low-rankness.

\paragraph{Low-rank via autocalibration\label{sec_LRviaautocalibration}}
We propose an approach where the decompression matrix $\rightMat$ is the product of two matrices $\autocalCompMat$ and $\dicCompMat$, i.e. $\rightMat = (\dicCompMat \autocalCompMat)^H$, and 
$\mathbf{X}_c = \compOp (\mathbf{X}) = \mathbf{X} \compMat = \mathbf{X} \dicCompMat \autocalCompMat  $.
The matrix  $\dicCompMat  \in \mathbb{C}^{\numTimePointsBeforCompression \times \dictCompValue}$ is the fingerprint dictionary compressed to its rank value $\dictCompValue$, while matrix $\autocalCompMat \in \mathbb{C}^{\dictCompValue \times \autocalliCompValue}$ is the result of an auto-calibration data and $\autocalliCompValue$ is the desired low-rank value. 
More formally, if $\dicMat \in \mathbb{C}^{\numPixels \times \numTimePointsBeforCompression}$ is the uncompressed dictionary matrix, it admits an SVD decomposition  $\dicMat = \mathbf{U}_d \mathbf{S}_d {\mathbf{V}_d}^H$, from which we obtain the dictionary SVD compression matrix $\dicCompMat$ by taking the $\dictCompValue$ first right singular vectors (i.e. the $\dictCompValue$ first columns of $\mathbf{V}_d$).
The auto-calibration data can be obtained directly in the compressed domain ($\mathbf{X}_{ac} = \compOp_d(\mathbf{X}_a) \in \mathbb{C}^{\numPixels \times \dictCompValue}$, with $\mathbf{X}_a \in \mathbb{C}^{\numPixels \times \numTimePointsBeforCompression}$), by for example running one iteration of AIR-MRF (i.e. one iteration of algorithm \ref{algo_IncGradProx} without the spatial regularization or similarly method \cite{mcgivney2014svd}), or compressed afterwards using the dictionary compression matrix $\dicCompMat$.
Note that in the first case, the data is in the Bloch manifold because a fingerprint matching (i.e. a Bloch manifold projection) has been performed as the last step of this algorithm. This is, however, different than using the dictionary for low-rank compression as in section \ref{sec_lowranksvd} because the distribution of the estimated tissues reflects what is (or rather what has been estimated to be) in the data as opposed to the fingerprint dictionary which is defined by a pre-chosen set of tissue parameters. In other words, the possible errors in this autocallibration only concern the distribussion of the tissues, but not the fingerprint signals themself, because the fingerprints are still coming from the dictionary.
We can then obtain the auto-calibration compression matrix $\autocalCompMat$ by performing an SVD decomposition on the compressed auto-calibration data $\compOp_d(\mathbf{X}_a) = \mathbf{X}_a \dicCompMat = \autocalCompU \autocalCompS {\autocalCompV}^H$, and keeping the $\autocalliCompValue$ first right singular vectors (i.e. the $\autocalliCompValue$ first columns of $\autocalCompV$).

The variants of our methods with auto-calibration are denoted \methodXXXVIII\ and \methodXXXIX\ in Table \ref{tab_proposedmthods}.
The variants with and without auto-calibration will be evaluated in section \ref{sec:csmrf_exp_brainweb}  (see methods \methodXXXIV\ vs  \methodXXXVIII\ and methods \methodXXXIX\ vs \methodXXIV\ in Table \ref{tab_proposedmthods} and Figure \ref{fig:brainweb_csmrf_noisy}).

% why autocalibr method
% show that it is a better low-rank approximation (Figure)

\subsection{Fingerprint matching}\label{sec_fingerMatch}

There are different techniques for fingerprint matching or more generally tissue parameters regression which are briefly discussed at the end of this section; however, the goal of this paper is not on the fingerprint matching technique, so we decided to rely on the classical matching technique \cite{ma2013magnetic} for this study.
The fingerprint matching consists of finding the most correlated fingerprint $\hat{k}_i$ in the dictionary for each voxel  $i$, i.e.
\begin{equation}\label{eq_ki}
\hat{k}_i = \argmax_{k} \frac{|  \langle \dicMat_{c(k,:)},  \hat{\mathbf{X}}_{c(i,:)}    \rangle  |}{ \| \dicMat_{c(k,:)} \|_2}.
\end{equation}
The proton density can then be estimated using the following equation:
\begin{equation}\label{eq_rho1}
\hat{\rho}_i = \max \left\{ \frac{ \real  \langle \dicMat_{c(\hat{k}_i,:)},  \hat{\mathbf{X}}_{c(i,:)}    \rangle  }{ \| \dicMat_{c(\hat{k}_i,:)} \|_2^2 }, 0 \right\},
\end{equation}
where $\real(\cdot)$ denotes the real part, $\dicMat_{c}$ is the discretized dictionary after compression ($\dicMat_{c} = \dicMat \dicCompMat $), and $ \dicMat_{c(k,:)}$ is the $k$-th fingerprint of the discretized dictionary.
Note that the fingerprint matching can be performed directly in the compressed domain since the dictionary decompression (adjoint of the compression operator) is an isometry 
(i.e. $ \langle \dicMat_{c(k,:)} \dicCompMat^H,  \hat{\mathbf{X}}_{c(i,:)} \dicCompMat^H   \rangle = \langle \dicMat_{c(k,:)},  \hat{\mathbf{X}}_{c(i,:)}    \rangle$, and $ \| \dicMat_{c(k,:)} \dicCompMat^H \|_2 = \| \dicMat_{c(k,:)} \|_2  $).

%(i.e. $  \langle \dicMat_{c(k,:)},  \hat{\mathbf{X}}_{c(i,:)}    \rangle = \langle \dicMat_{c(k,:)} \dicCompMat^H,  \hat{\mathbf{X}}_{c(i,:)} \dicCompMat^H   \rangle =  \langle \dicMat_{(k,:)},  \hat{\mathbf{X}}_{(i,:)}    \rangle$, and $ \| \dicMat_{c(k,:)} \|_2 = \| \dicMat_{(k,:)} \dicCompMat^H \|_2 = \| \dicMat_{(k,:)} \|_2   $).

In the case we use the low-rank variant of the methods with auto-calibration data (described in section \ref{sec_LRviaautocalibration}, i.e. methods \methodXXXVIII\ and \methodXXXIX), the previous formulas \eqref{eq_ki} and \eqref{eq_rho1} need to take into account the auto-calibration compression matrix.
As $ \langle \dicMat_{c(k,:)},  \hat{\mathbf{X}}_{c(i,:)} \autocalCompMat^H   \rangle = \langle \dicMat_{c(k,:)} \autocalCompMat,  \hat{\mathbf{X}}_{c(i,:)}   \rangle$, we actually don't need to uncompress the data $\hat{\mathbf{X}}_{c(i,:)}$ with $\autocalCompMat^H$ prior to the fingerprint matching and can instead apply (only once) the compression operator $\autocalCompMat$ on the dictionary.
So that with auto-calibration, formulas \eqref{eq_ki} and \eqref{eq_rho1} need then to be replaced by:
$$\hat{k}_i = \argmax_{k} \frac{|  \langle \dicMat_{c(k,:)} \autocalCompMat,  \hat{\mathbf{X}}_{c(i,:)}    \rangle  |}{ \| \dicMat_{c(k,:)} \|_2},$$
\begin{equation}\label{eq_rho2}
\hat{\rho}_i = \max \left\{ \frac{ \real  \langle \dicMat_{c(\hat{k}_i,:)} \autocalCompMat,  \hat{\mathbf{X}}_{c(i,:)}    \rangle  }{ \| \dicMat_{c(\hat{k}_i,:)} \|_2^2 }, 0 \right\}
\end{equation}

Once the fingerprint $\hat{k}_i$ is identified, the quantitative tissue values $\hat{\theta}_i = \{ \text{T}1_i, \text{T}2_i, \ldots \}$  are obtained by a simple look-up table (LUT): $\hat{\theta}_i = \text{LUT}(\hat{k}_i)$,
while the proton density is obtained by equations \eqref{eq_rho1} (if no auto-calibration) or \eqref{eq_rho2} (if auto-calibration).
The contrast image is then re-synthesized by multiplying the proton density with the selected atom as follows: $\hat{\mathbf{X}}_{c(i,:)}^{new} := \hat{\rho}_i \dicMat_{c(\hat{k}_i ,:)},$
and including the auto-calibration compression in the case of the auto-calibration variant: $\hat{\mathbf{X}}_{c(i,:)}^{new} := \hat{\rho}_i \dicMat_{c(\hat{k}_i ,:)} \autocalCompMat.$

Rather than selecting the most correlated fingerprint by exhaustive search, fingerprint matching can be accelerated with approximate k-nearest-neighbour (ANN) techniques as in AIR-MRF \cite{cline2017air} where a kd-tree is used.
Other fast matching techniques have been developed specially for MRF such as the Fast Group Matching (FGM) \cite{cauley2015fast} or neural networks \cite{cohen2017deep,fang2019deep,song2019magnetic}. However, the choice of the accelerated matching technique is beyond the scope of this paper and only exhaustive search was used for all the reconstruction methods tested in section \ref{sec_experiments}.

\subsection{Spatial regularization}

Spatial regularization (term $\regterm(\hat{\mathbf{X}}_c)$ in Eq. \eqref{eq_optimProblem}) is applied via its proximal operator \cite{combettes2011proximal} in the compressed domain, on each image slice of $\hat{\mathbf{X}}_c$.
Because the low-rank value $\autocalliCompValue$ is small (in the order of $10$) compared to the sequence length $\numTimePointsBeforCompression$ (in the order of $10^3$), the regularization is quite efficient.
We choose to use an isotropic TV regularization using Chambolle algorithm \cite{chambolle2004algorithm}, because it converges very fast in the first iterations, and thus provides a good approximation of the TV proximal operation after just a few iterations. 
Other efficient spatial regularization such as cycle spinning \cite{kamilov2014variational,mailhe2018fast} can be used instead. Joint (along the compressed-temporal dimension) TV sparse regularization also provides similar performance in our experience.

\subsection{Step-size selection\label{sec:stepSizeSel}}

Selection of the step size is important to ensure convergence and good reconstruction performance.
Forward backward algorithm is known to converge for $\alpha < 2/L(\nabla f )$, where $L(\nabla f)$ denotes the Lipschitz constant of the gradient of $f$.
In our case, $f = \frac{1}{2} \| \samplOp ( \hat{\mathbf{X}}_c ) - \mathbf{Y}  \|_2^2$ and thus $L(\nabla f )$ is simply the spectral radius of $\samplOp^H \samplOp$ which should be equal to one if the (non-uniform) Fourier transform and sampling operations were properly normalized and compensated.
In practice, one rarely has an accurate knowledge of $L(\nabla f )$ and also this condition is conservative in that it focuses on the worst case.
% guaranties convergence for all the cases including the worst case scenario. In other words it may be quite conservative, i.e. leading to a small step size and a slow converge, for an average case scenario.
It is thus usually better in practice to start with a relatively large step size to converge quickly but at the same time ensure stability of the iterations with a backtracking rule as  explained in section \ref{sec:backtracking}. The drawback of a backtracking rule, however, is that it usually increases significantly the computation load of the algorithm.

Previous work on step-size selection in the context of iterative MRF reconstruction, such as in BLIP \cite{davies2014compressed} and AIR-MRF \cite{cline2017air}, suggested that a large step size (larger than $\alpha=1$) ensures faster convergence and leads to significantly better results.
In \cite{davies2014compressed}, the proposed heuristic is to start with a step size equals to the undersampling ratio e.g. $\alpha = \numPixels/\numSamplesPerRO$ and then to use a backtracking rule similar to the one presented in section \ref{sec:backtracking}.

\subsubsection{Rescaling step-size\label{sec:rescaling}}
In AIR-MRF \cite{cline2017air}, if $\alpha > 0$ is the original step size, the first iteration of AIR-MRF leads to $\hat{\mathbf{X}}_c^{(1)} = \dictProj(\alpha \samplOp^H  \mathbf{Y})  = \alpha \dictProj( \samplOp^H  \mathbf{Y})$ because of the  positive homogeneity property of the orthogonal projection onto a cone (the Bloch manifold). 
As a result, the step size $\alpha$ of the first iteration can be chosen at the end of the first iteration, after the matching step. Moreover, it can be chosen optimaly by minimizing the data fidelity term: $ \| \alpha \samplOp ( \widetilde{\mathbf{X}}_c^{(1)} ) -  \mathbf{Y}\|_2^2$ with $\widetilde{\mathbf{X}}_c^{(1)}  = \dictProj(\samplOp^H  \mathbf{Y})$, where  closed form solution is $\alpha =  \real \left( \langle  \mathbf{Y}, \samplOp ( \widetilde{\mathbf{X}}_c^{(1)} ) \rangle \right)  \big/  \| \samplOp ( \widetilde{\mathbf{X}}_c^{(1)}) \|_2^2  $.
We call this procedure which is used in all our proposed methods (see Table \ref{tab_proposedmthods}): \textit{rescaling step-size} and use it as an heuristic to select the step size at the first iteration and subsequent iterations until a backtracking rule is possibly applied.

\subsubsection{Backtracking strategy\label{sec:backtracking}}

Convergence can be guaranteed via a backtracking line search. In general, such method proceeds by first checking a condition at the end of each iteration. 
The so-called backtracking condition enforces that the objective has decreased ``sufficiently''. If this condition fails, the stepsize is decreased and the current iteration is restarted until the backtracking condition holds.
The backtracking rule described here is the same in AIR-MRF and inspired by the one of IHT \cite{blumensath2010normalized,davies2014compressed,cline2017air}.
The backtracking condition is that if either \textit{condition a} or \textit{condition b} defined below is false, then the candidate update is rejected, and the iteration is restarted with the step size halved.

\textit{condition a}:
$$\alpha \leq 0.99 \frac{ \left\| \hat{\mathbf{X}}_c^{(\iter+1) } -  \hat{\mathbf{X}}_c^{(\iter) }\right\|_2^2  }{\left\| \samplOp^H \samplOp  \left( \hat{\mathbf{X}}_c^{(\iter+1) } -  \hat{\mathbf{X}}_c^{(\iter)} \right)  \right\|_2^2} $$

\textit{condition b}:
$$ \left\|   \samplOp^H  \mathbf{Y} - \samplOp^H \samplOp \hat{\mathbf{X}}_c^{(\iter+1) } \right\|_2^2  \geq   \left\|   \samplOp^H  \mathbf{Y} - \samplOp^H \samplOp \hat{\mathbf{X}}_c^{(\iter) } \right\|_2^2 $$

The iteration is repeated until both conditions are met.
In the following, we call \textit{backtracking} (BT) the strategy consisting of rescaling the step size (as described in \ref{sec:rescaling}) in the first iteration and then applying the backtracking rule just described.

\subsubsection{Rescaling fixed step-size\label{sec:rescaling_fixed_step_size}}

We call \textit{Rescaling fixed step-size} strategy (FSZ) the strategy consisting of rescaling the step size (as described in \ref{sec:rescaling}) and then keeping the step size fixed for the remaining iterations.

These step-size strategies will be evaluated in the next section where variants of the same method with FSZ or with BT strategy will be compared (see e.g. methods \methodXXIV\ and \methodXXIV BT in  Table \ref{tab_proposedmthods}).

%\subsubsection{cheap step-size strategies}
%As the backtracking strategy increases the computational load of the CS-MRF algorithms, we propose here some heuristics to select the step size with a negligeable complexity. 
%One first approach is to select a fixed step-size value that is a fraction of the rescaling step-size of section \ref{sec:rescaling_fixed_step_size}.
%Another approach is to schedule a decreasing step size ones the image consistency error stops making progress.
%The decreasing step size proposed here follows a $1/(\iter-\iter_0)$ rule (where $\iter$ is the iteration index) so that its limit is zero but it sum is unbouded in order to not decrease too fast (see \cite{bertsekas1999nonlinear} for a justification of such step-size rules in the case of gradient or gradient proximal algorithms). 

\section{Experiments\label{sec_experiments}}

% we use plotReconstructedImage.m to plot the images
% batchfig2png.m was used to process the .fig to .png

% tab of the methods
\begin{table}[]
\caption{Proposed methods}
\label{tab_proposedmthods}
\def\arraystretch{1.1}%  1 is the default, change whatever you need
\begin{tabular}{ l|l|l|l|l }
\shortstack{name of\\the method}
  & \multicolumn{1}{c|}{\shortstack{optimization\\ algorithm}} & \shortstack{auto-\\calibration} &  \shortstack{matching at\\every iteration} & \shortstack{step-size\\ strategy} \\
  \hline
\methodXXIV          	& algo \ref{algo_IncGradProx}   	& no             	& yes		& FSZ                        \\
\methodXXIV BT      	& algo \ref{algo_IncGradProx}   	& no             	& yes		& BT                        \\
\methodXXXIV         	& algo \ref{algo_IncGradProx}   	& no              & no		& BT                          \\
\methodXXXVIII       & algo \ref{algo_IncGradProx}   	& yes            	& no		& BT                          \\
\methodXXXIX        	& algo \ref{algo_IncGradProx}   	& yes           	& yes		& FSZ                         \\
\methodGFB         	& algo  \ref{algo_GFB}  		& no              & yes		& BT                        
\end{tabular}
\end{table}

We studied the different MRF algorithms on the Brainweb digital phantom data \cite{cocosco1997brainweb} %for different sequence lengths
 in order to perform a quantitative analysis of the performance. Brainweb digital phantom being the only data for which we have a ground truth.
We then showed results on real acquisition, on the NIST phantom in section \ref{sec_NIST} and on a real volunteer brain data in section \ref{sec_volunteer}.
We studied the performance when the raw measurements are noise-free, and when the measurements were contaminated with some noise in order to make the experiment more realistic and study the robustness of the algorithms.
We compared in section \ref{sec:csmrf_exp_brainweb} our proposed methods and their variants (summarized in Table \ref{tab_proposedmthods}) which perform spatial regularization and showed that  \methodGFB\ is the best method. We also compared \methodGFB\ with other methods such as the classical MRF \cite{ma2013magnetic}, AIR-MRF \cite{cline2017air}, kt-SVD-MRF \cite{ktsvdmrf2017}, and Zhao et al. LR \cite{zhao2017improved}. %\ref{sec:soa_exp}.
In all the cases, the results of our proposed methods were shown for the iteration which obtained the lowest data fidelity error $\| \mathbf{Y} -  \samplOp ( \hat{\mathbf{X}}_c ) \|_2^2$, which is usually obtained after 5 to 10 iterations, and a SVD compression of $\autocalliCompValue=10$ as explained in section \ref{sec_lowranksvd}.

The performance is measured in terms of relative error in the tissue map estimation, i.e. if $\hat{\theta}_n$ is the estimated tissue parameter (T1 or T2) map at pixel location $n$, and $\theta_n$ is the ground truth tissue parameter map at the same pixel location, then the error is calculated as: $ \frac{1}{N} \sum_{n} | \hat{\theta}_n -  \theta_n | / \theta_n$. Note that a mask is applied so that the background is discarded in the error calculation.

\subsection{MRF Acquisitions\label{sec_sequence}}

%All the experiments of this paper are obtained using the following pulse sequence \cite{cline2017air}.
%A prototype unbalanced SSFP pulse sequence with an initial inversion pulse followed by a train of pseudorandom flip angles (FAs, 0-75\arcdeg) and repetition times (TRs, 12-15ms) is used \cite{jiang2015mr}. 
%A dual-density spiral trajectory is designed with 48 interleaves \cite{hargreaves2005}, inner/outer density (2,1) transitioning at a relative radius of 0.45. 
%The spiral readout duration is 5.0 ms sampled with 2.5 \micro s dwell time and TE of 3.7 ms. 
%At each TR k-space data is collected on one of the 48 spiral interleaves. 
%The single-shot spiral interleaf is uniformly rotated by 7.5\arcdeg\ with each TR. 

All the experiments of this paper (both the synthetic on Brainweb, and the real scans) were obtained using the following prototype pulse sequence: 
After the application of an inversion pulse and an inversion time (TI) of 21 ms the FISP acquisition started. Here, the flip angle (FA) and the repetition time (TR) varied for each
echo. TR ranged from 12.1 ms to 15.0 ms and the FA ranged from 0\arcdeg\ to 74\arcdeg\ while the echo time was constant (2 ms). 
Each echo encoded an image by a single spiral readout with a variable density k-space
trajectory as described in \cite{meyer2011dual}, using an inner and outer undersampling ratio of 24 and 48 respectively.
The spiral readout had approx. 6 ms duration for the default parameters (FOV 256, matrix
256, dwell time 2.5 \micro s, 48 spiral interleaves for full sampling) and was followed by gradients rewinding the
moments on the x- and y-axis.
In order to improve the incoherence of resulting undersampling artefacts, the spiral
trajectory was rotated by 82.5\arcdeg\ (84\arcdeg\ every 48 interleaves) from TR to TR \cite{pfeuffer2017mitigation}.

A long sequence of length 3000 TRs was acquired and used as a reference to compare the different methods which were evaluated on shorter sequences (e.g. of length 1000 TRs for the real acquisition experiments, and from 600 to 2000 TRs for the Brainweb experiment) obtained by retrospectively discarding data from the end of the original sequence.
%Acquisitions at shorter sequence lengths were retrospectively generated by discarding data from the end of the sequence.
The constructed dictionary consisted of 5366 atoms with T1 ranging from 10 to 4500 ms (increments of 10, 20, 40, and 100 for ranges 10 to 100, 120 to 1000, 1040 to 2000, and 2050 to 4500, respectively, in ms) and T2 ranging from 2 to 3000 ms (increments of 2, 5, 10, 50, 100, and 200 for ranges 2 to 10, 15 to 100, 110 to 300, 350 to 800, 900 to 1600, and 1800 to 3000, respectively, in ms), excluding combinations of T1 and T2 values where T1 $<$ T2.

\subsection{Brainweb digital phantom evaluations\label{sec:csmrf_exp_brainweb}}

In the following experiments, the digital phantom from Brainweb \cite{cocosco1997brainweb} was used with ground truth T1, T2 and proton density (PD) maps at a field strength of 1.5 T, zero padded
to be square.  The 434 $\times$  434 image is nearest-neighbor downsampled to 256 $\times$ 256. Bloch responses for both dictionary generation and generation of simulated data were calculated from ground truth values using Bloch simulations. 
Note that as described in section \ref{sec_sequence} only one spiral is acquired per time point. The (simulated) data in this section is single coil while the real scan acquisitions in the next sections are multi-coils.
The methods are evaluated for different sequence lengths (by retrospectively discarding the end of the original sequence) ranging from 600 TRs to 3000 TRs.
% Data both with and without noise were used for reconstructions.

\subsubsection{Noise-free measurements\label{sec:noiselesscase}}
% Brainweb without noise, different sequence length
% NIST different sequence length
% Volunteer different sequence length

We first studied the performance of our methods with respect to the sequence length on noise-free Brainweb digital phantom data. 
Results (not reported) showed that our proposed methods (summarized in table \ref{tab_proposedmthods}) performed quite similarly in the noise-free case.
%Results reported in tables \ref{tab_csmrfBrainwebT1_noiseless} and \ref{tab_csmrfBrainwebT2_noiseless} in appendix (section \ref{sec:app_csmrf_noiseless}) show that our proposed methods (summarized in table \ref{tab_proposedmthods}) perform quite similarly in the noise-free case.
%Volunteer brain MR data was acquired on a MAGNETON Aera 1.5T MR scanner (Siemens Healthcare, Erlangen, Germany). 
The pulse sequence and dictionary used is described in section \ref{sec_sequence}.

The regularization parameter was set to $\lambda=10^{-4}$ for all the methods, which is the optimal value we obtained by swiping this parameter value.
The reasons we chose a fix value for all the sequence lenghs and variants are:  1) these algorithms solve the same optimization problem (Equation \eqref{eq_optimProblem}), and 2) we want to see if a method is more robust or sensitive to this parameter when the conditions of the experiment change slightly (the sequence length is changing but the SVD compression rank $k$ remains the same). When we will add a little noise in section \ref{sec:noisycase} this parameter will be set slightly higher as this parameter should be proportional to the noise std \cite{chen2001atomic}.

\subsubsection{Influence of noise\label{sec:noisycase}}

\begin{figure}
  \begin{subfigure}[b]{\columnwidth}
    \includegraphics[width=\textwidth]{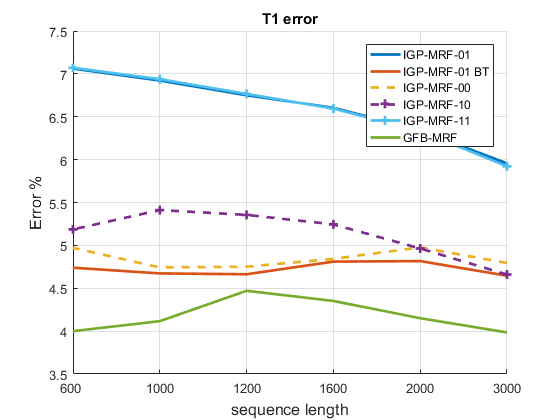}
    %\caption{T1 error}
    \label{fig:brainwebT1_csmrf_noisy}
  \end{subfigure}
  \begin{subfigure}[b]{\columnwidth}
    \includegraphics[width=\textwidth]{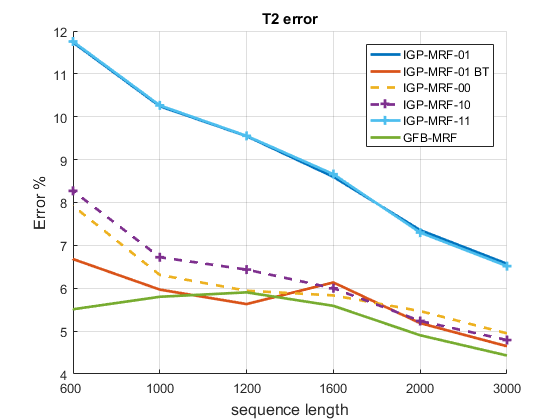}
    %\caption{T2 error}
    \label{fig:brainwebT2_csmrf_noisy}
   \end{subfigure}
   \caption{T1 (top) and T2 (bottom) errors w.r.t. sequence length of proposed methods on Brainweb digital phantom data with noisy measurements}
   \label{fig:brainweb_csmrf_noisy}
\end{figure}

In this experiment (see Figure \ref{fig:brainweb_csmrf_noisy}), we want to assess the robustness of the methods with respect to noise in the measurements.
We followed the same experimental setup as in section \ref{sec:noiselesscase}, but with a (zero mean) Gaussian noise added to the raw measurements.
The standard deviation of the noise was set to be equal to 0.1\% of the value of the largest sample of the raw measurement.
The regularization parameter was set to $\lambda=5\times 10^{-4}$ for all the methods, which is the optimal value we obtain by swiping this parameter value.

%\begin{figure}
%  \includegraphics[width=\linewidth]{figures/CSMRF_brainwebT1_noise.png}
%  \caption{T1 error of the proposed CS-MRF methods with noisy measurements for different sequence length}
%  \label{fig:brainwebT1_csmrf_noisy}
%\end{figure}
%
%\begin{figure}
%  \includegraphics[width=\linewidth]{figures/CSMRF_brainwebT2_noise.png}
%  \caption{T2 error of the proposed CS-MRF methods with noisy measurements for different sequence length}
%  \label{fig:brainwebT2_csmrf_noisy}
%\end{figure}

%\subsubsection{Results}

%The different methods obtained similar results in the noiseless case (results in the noiseless case are depicted in Appendix in tables \ref{tab_csmrfBrainwebT1_noiseless} and \ref{tab_csmrfBrainwebT2_noiseless}). In the noisy case however, \methodGFB\ obtained the best results in all the cases  (see Figures \ref{fig:brainwebT1_csmrf_noisy} and \ref{fig:brainwebT2_csmrf_noisy}).

Results with noisy measurements, depicted in Figure \ref{fig:brainweb_csmrf_noisy},  show that \methodGFB\ obtained the best results on both T1 and T2 map estimation for (almost) all the sequence lengths.

% except for T1 estimation in the noiseless case (Figure \ref{fig:brainwebT1_csmrf_noiseless}) where it obtained the best results after (and with only $0.1 \%$ difference in the error) \methodXXIV\ with constant large step-size strategy and \methodXXXIX. However, these last two methods performed poorly when the measurements are contaminated with noise (see Figures \ref{fig:brainwebT1_csmrf_noisy} and \ref{fig:brainwebT2_csmrf_noisy}).
%So overall \methodGFB\ is the method which obtained the best results.

%\subsection{Comparison with the state of the art\label{sec:soa_exp}}

We compared \methodGFB\ which obtained globally the best results in the previous experiment (see Figure \ref{fig:brainweb_csmrf_noisy}) with the classical MRF \cite{ma2013magnetic}, AIR-MRF \cite{cline2017air}, kt-SVD-MRF \cite{ktsvdmrf2017}, and Zhao et al. LR \cite{zhao2017improved}. For the last method, we set the low-rank value to $k=8$ as suggested by the authors \cite{zhao2017improved}. Note that we also tested Zhao et al. method with $k=10$ in order to select the same rank thresholding value as for our methods and the results were very similar (but slightly worse) as with $k=8$.

%\subsubsection{Noiseless measurements\label{sec:soa_noiselesscase}}

\begin{figure}
  \begin{subfigure}[b]{\columnwidth}
    \includegraphics[width=\textwidth]{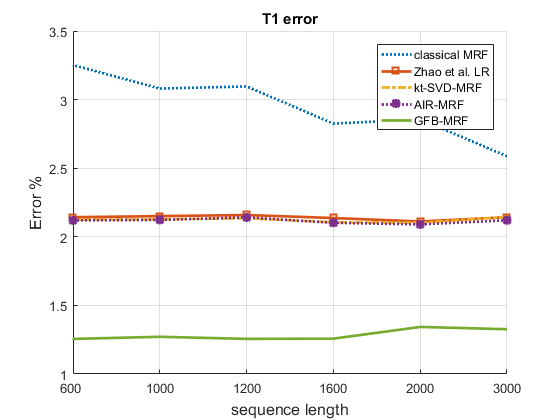}
  \end{subfigure}
  \begin{subfigure}[b]{\columnwidth}
    \includegraphics[width=\textwidth]{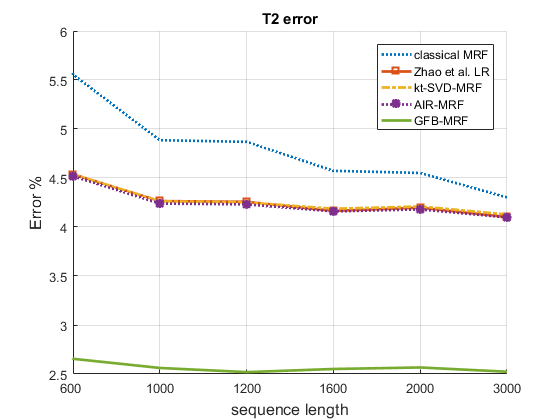}
   \end{subfigure}
   \caption{T1 (top) and T2 (bottom) errors w.r.t. sequence length of proposed \methodGFB\  method compared to other methods on Brainweb digital phantom data with noise-free measurements}
   \label{fig:brainwebsoa_noiseless}
\end{figure}

\begin{figure}
  \begin{subfigure}[b]{\columnwidth}
    \includegraphics[width=\textwidth]{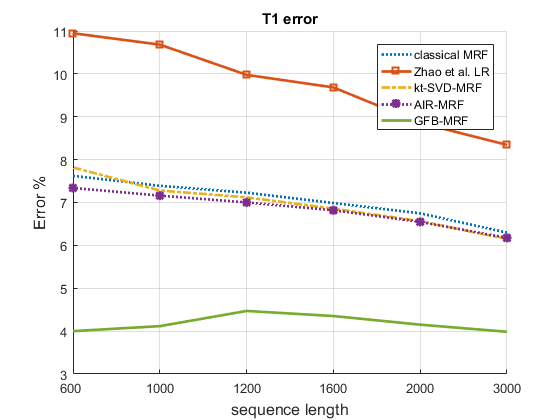}
  \end{subfigure}
  \begin{subfigure}[b]{\columnwidth}
    \includegraphics[width=\textwidth]{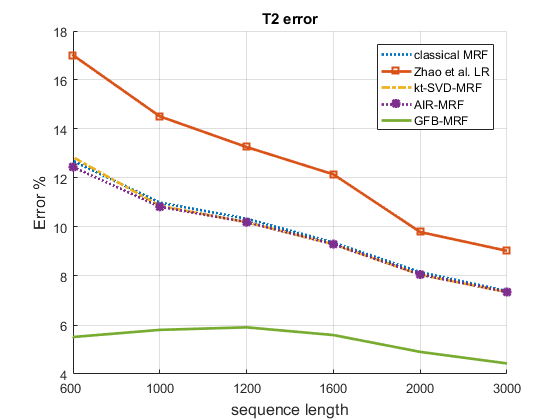}
   \end{subfigure}
   \caption{T1 (top) and T2 (bottom) errors w.r.t. sequence length of proposed \methodGFB\  method compared to the other methods on Brainweb digital phantom data with noisy measurements}
   \label{fig:brainwebsoa_noisy}
\end{figure}

The performance of the different algorithms on noise-free Brainweb digital phantom data are depicted in Figure \ref{fig:brainwebsoa_noiseless}, and the performance of the different algorithms on Brainweb digital phantom data with noisy measurements are depicted in Figure \ref{fig:brainwebsoa_noisy}.

%\begin{figure}
%  \includegraphics[width=\linewidth]{figures/SOA_brainwebT1.png}
%  \caption{T1 error of our proposed \methodGFB\  method with respect to the state of the art on noise-free brainweb image}
%  \label{fig:brainwebT1_soa_noiseless}
%\end{figure}
%
%\begin{figure}
%  \includegraphics[width=\linewidth]{figures/SOA_brainwebT2.png}
%  \caption{T2 error of our proposed \methodGFB\  method with respect to the state of the art on noiseless brainweb image}
%  \label{fig:brainwebT2soa_noiseless}
%\end{figure}

%\subsubsection{Influence of noise}

%The performance of the different algorithms on Brainweb with noisy measurements are depicted in Figures \ref{fig:brainwebT1_soa_noisy} and \ref{fig:brainwebT2soa_noisy}.

%previously established methods

%\begin{figure}
%  \includegraphics[width=\linewidth]{figures/SOA_brainwebT1_noise.png}
%  \caption{T1 error of our proposed  \methodGFB\ method with respect to the state of the art with noisy measurements on  brainweb image}
%  \label{fig:brainwebT1_soa_noisy}
%\end{figure}
%
%\begin{figure}
%  \includegraphics[width=\linewidth]{figures/SOA_brainwebT2_noise.png}
%  \caption{T2 error of our proposed \methodGFB\  method with respect to the state of the art with noisy measurements on  brainweb image}
%  \label{fig:brainwebT2soa_noisy}
%\end{figure}

% more results in a tab: PD, MSE

%\subsubsection{Results}

\methodGFB\ performed much better than the classical MRF method and the other compared methods (MRF, AIR-MRF, \ZhaoLR, kt-SVD-MRF) in both noise-free and noisy case and for all the sequence lengths.
As an example, \methodGFB\ reduced the T1 error from 2.1\% to 1.3\% and the T2 error from 4.5\% to 2.7\% compared to AIR-MRF, \ZhaoLR, and kt-SVD-MRF at a sequence length of L=600 in the noiseless case,
and the T1 error from 7.5\% to 4\% and the T2 error from 12.5\% to 5.5\% compared to the classical MRF, AIR-MRF, and kt-SVD-MRF at a sequence length of L=600 in the added noise case.
We also noted that AIR-MRF, \ZhaoLR\ and kt-SVD-MRF perform significantly better than the original MRF method in the noise-free case, but that in the noisy scenario they all performed similarly as the classical MRF except \ZhaoLR\ \cite{zhao2017improved} which performed worse.

\subsection{NIST phantom evaluations\label{sec_NIST}}

\begin{figure}[htb]
\centering
  \rotatebox[origin=l]{90}{\makebox[1in]{classical MRF}}%
  \hspace{0pt}
  \begin{subfigure}[b]{.32\linewidth}
    \centering
    \includegraphics[width=.99\textwidth]{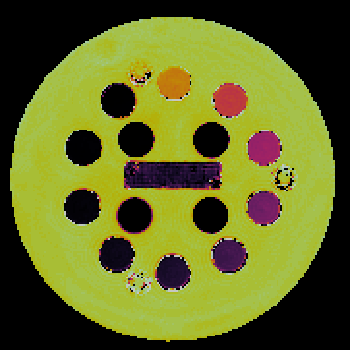}
  \end{subfigure}%   
  \begin{subfigure}[b]{.32\linewidth}
    \centering
    \includegraphics[width=.99\textwidth]{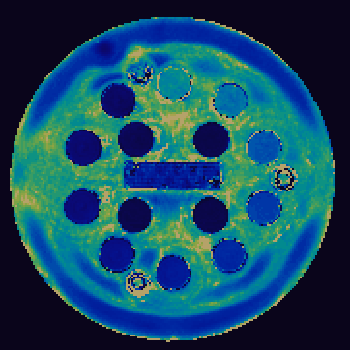}
  \end{subfigure}%  
    \begin{subfigure}[b]{.32\linewidth}
    \centering
    \includegraphics[width=.99\textwidth]{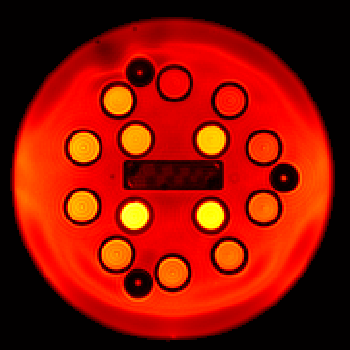}
  \end{subfigure}\\%
    \rotatebox[origin=l]{90}{\makebox[1in]{AIR-MRF}}%
    \hspace{0pt}
    \begin{subfigure}[b]{.32\linewidth}
    \centering
    \includegraphics[width=.99\textwidth]{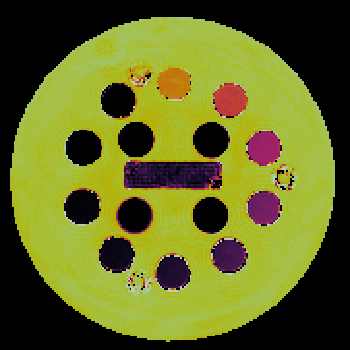}
  \end{subfigure}%   
  \begin{subfigure}[b]{.32\linewidth}
    \centering
    \includegraphics[width=.99\textwidth]{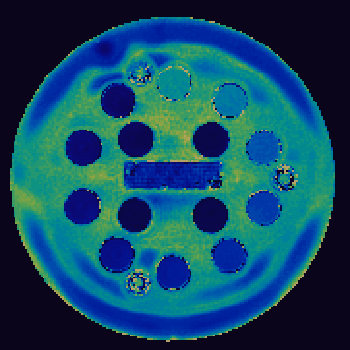}
  \end{subfigure}%  
    \begin{subfigure}[b]{.32\linewidth}
    \centering
    \includegraphics[width=.99\textwidth]{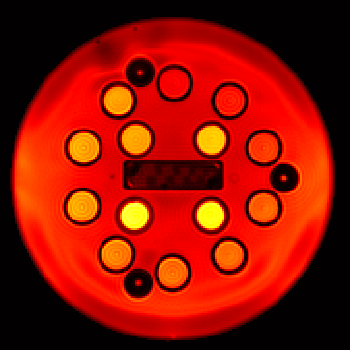}
  \end{subfigure}\\%
    \rotatebox[origin=l]{90}{\makebox[1in]{kt-SVD-MRF}}%
  \hspace{0pt}
   \begin{subfigure}[b]{.32\linewidth}
    \centering
    \includegraphics[width=.99\textwidth]{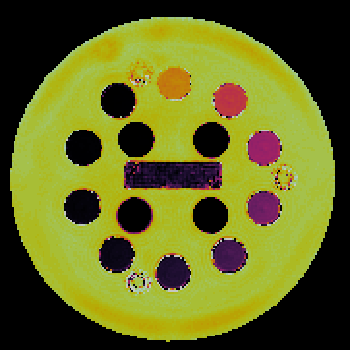}
  \end{subfigure}%   
  \begin{subfigure}[b]{.32\linewidth}
    \centering
    \includegraphics[width=.99\textwidth]{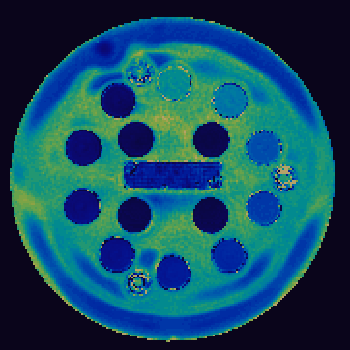}
  \end{subfigure}%  
    \begin{subfigure}[b]{.32\linewidth}
    \centering
    \includegraphics[width=.99\textwidth]{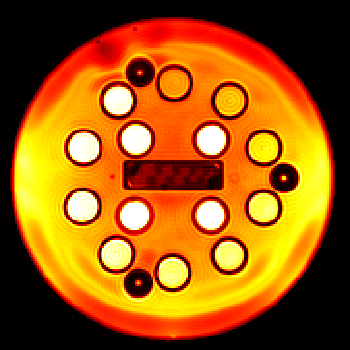}
  \end{subfigure}\\%
    \rotatebox[origin=l]{90}{\makebox[1.in]{\methodGFB}}%
  \hspace{0pt}
      \begin{subfigure}[b]{.32\linewidth}
    \centering
    \includegraphics[width=.99\textwidth]{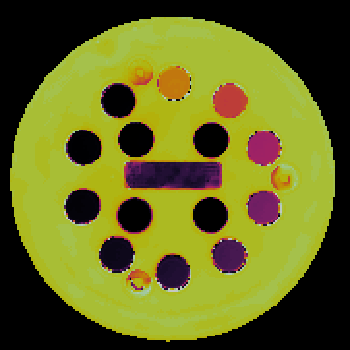}
  \end{subfigure}%   
  \begin{subfigure}[b]{.32\linewidth}
    \centering
    \includegraphics[width=.99\textwidth]{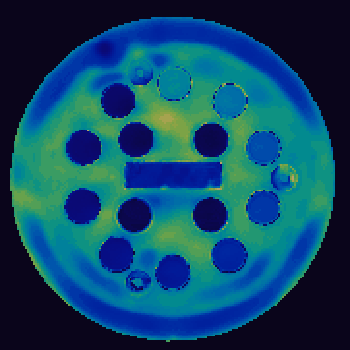}
  \end{subfigure}%  
    \begin{subfigure}[b]{.32\linewidth}
    \centering
    \includegraphics[width=.99\textwidth]{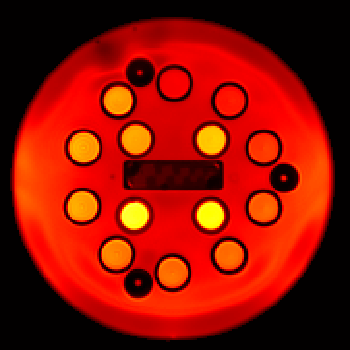}
  \end{subfigure}\\%
      \rotatebox[origin=l]{90}{\makebox[.5in]{\quad}}%
  \hspace{0pt}
  \begin{subfigure}[b]{.32\linewidth}
    \centering
    \includegraphics[width=.99\textwidth]{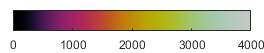}
    \caption{T1}\label{fig:1a}
  \end{subfigure}%   
  \begin{subfigure}[b]{.32\linewidth}
    \centering
    \includegraphics[width=.99\textwidth]{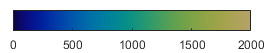}
    \caption{T2}\label{fig:1b}
  \end{subfigure}%  
    \begin{subfigure}[b]{.32\linewidth}
    \centering
    \includegraphics[width=.99\textwidth]{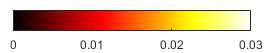}
    \caption{PD}\label{fig:1b}
  \end{subfigure}%
  \caption{NIST phantom evaluations with noise-free measurements and sequence length 1000}\label{fig_NIST_nonoise}
\end{figure}

\begin{figure}[htb]
\centering
  \rotatebox[origin=l]{90}{\makebox[1in]{classical MRF}}%
  \hspace{0pt}
  \begin{subfigure}[b]{.32\linewidth}
    \centering
    \includegraphics[width=.99\textwidth]{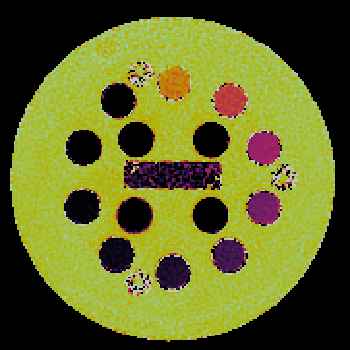}
  \end{subfigure}%   
  \begin{subfigure}[b]{.32\linewidth}
    \centering
    \includegraphics[width=.99\textwidth]{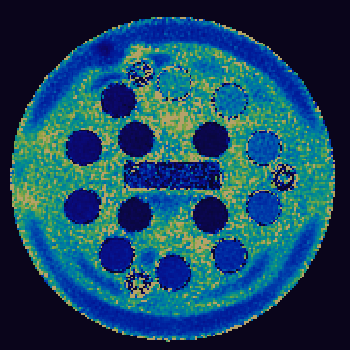}
  \end{subfigure}%  
    \begin{subfigure}[b]{.32\linewidth}
    \centering
    \includegraphics[width=.99\textwidth]{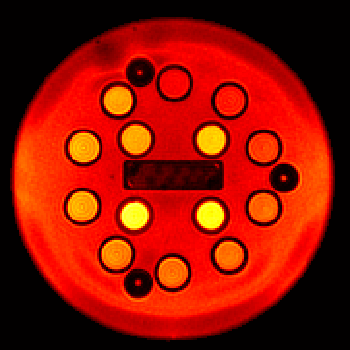}
  \end{subfigure}\\%
    \rotatebox[origin=l]{90}{\makebox[1in]{AIR-MRF}}%
  \hspace{0pt}
    \begin{subfigure}[b]{.32\linewidth}
    \centering
    \includegraphics[width=.99\textwidth]{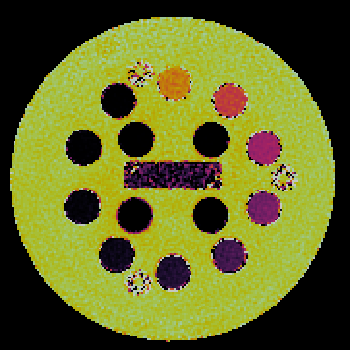}
  \end{subfigure}%   
  \begin{subfigure}[b]{.32\linewidth}
    \centering
    \includegraphics[width=.99\textwidth]{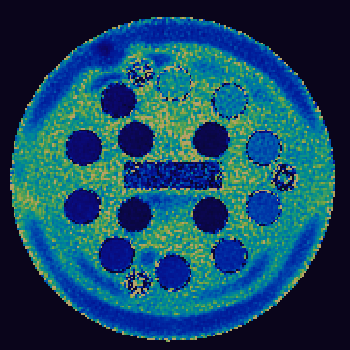}
  \end{subfigure}%  
    \begin{subfigure}[b]{.32\linewidth}
    \centering
    \includegraphics[width=.99\textwidth]{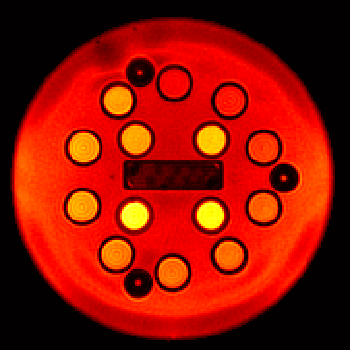}
  \end{subfigure}\\%
    \rotatebox[origin=l]{90}{\makebox[1in]{ kt-SVD-MRF}}%
  \hspace{0pt}
   \begin{subfigure}[b]{.32\linewidth}
    \centering
    \includegraphics[width=.99\textwidth]{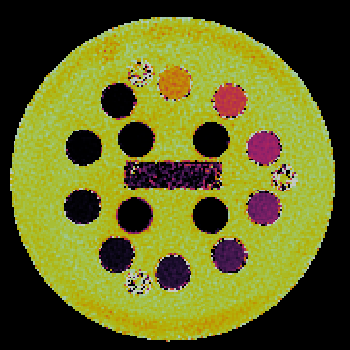}
  \end{subfigure}%   
  \begin{subfigure}[b]{.32\linewidth}
    \centering
    \includegraphics[width=.99\textwidth]{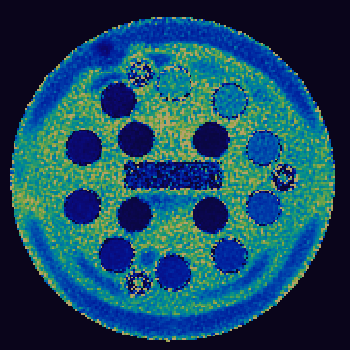}
  \end{subfigure}%  
    \begin{subfigure}[b]{.32\linewidth}
    \centering
    \includegraphics[width=.99\textwidth]{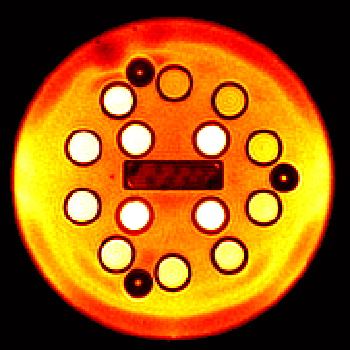}
  \end{subfigure}\\%
    \rotatebox[origin=l]{90}{\makebox[1in]{\methodGFB}}%
  \hspace{0pt}
  \begin{subfigure}[b]{.32\linewidth}
    \centering
    \includegraphics[width=.99\textwidth]{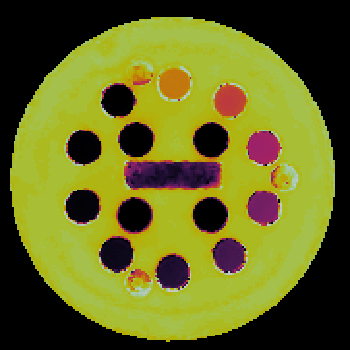}
  \end{subfigure}%   
  \begin{subfigure}[b]{.32\linewidth}
    \centering
    \includegraphics[width=.99\textwidth]{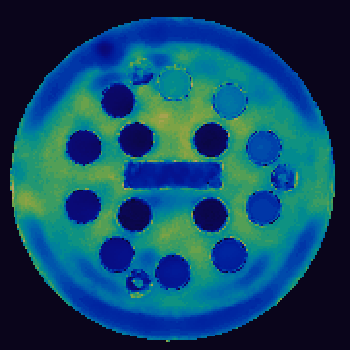}
  \end{subfigure}%  
    \begin{subfigure}[b]{.32\linewidth}
    \centering
    \includegraphics[width=.99\textwidth]{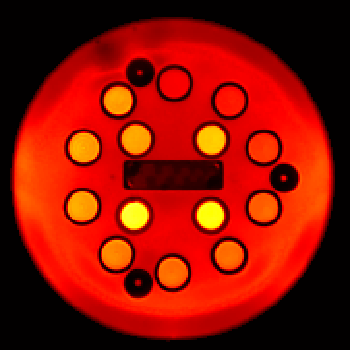}
  \end{subfigure}\\%
        \rotatebox[origin=l]{90}{\makebox[.5in]{\quad}}%
  \hspace{0pt}
  \begin{subfigure}[b]{.32\linewidth}
    \centering
    \includegraphics[width=.99\textwidth]{figures/colormapT1Nist.png}
    \caption{T1}\label{fig:1a}
  \end{subfigure}%   
  \begin{subfigure}[b]{.32\linewidth}
    \centering
    \includegraphics[width=.99\textwidth]{figures/colormapT2Nist.png}
    \caption{T2}\label{fig:1b}
  \end{subfigure}%  
    \begin{subfigure}[b]{.32\linewidth}
    \centering
    \includegraphics[width=.99\textwidth]{figures/colormapPDNist.png}
    \caption{PD}\label{fig:1b}
  \end{subfigure}%
  \caption{NIST phantom evaluations with noisy measurements and sequence length 1000}\label{fig_NIST_with_noise}
\end{figure}

In this section, we studied the performance of our proposed \methodGFB\ method and other competitive methods on the NIST phantom \cite{nistphantom}, 1) in the noise-free case (Figure \ref{fig_NIST_nonoise}), and 2) with additional Gaussian noise added to the measurements as described in section \ref{sec:noisycase} (Figure \ref{fig_NIST_with_noise}). Zoom on these images are also depicted in the supporting document.
We used the colormap specifically defined for MRF  in \cite{griswoldMRFcolor2018}.
NIST phantom was scanned using the sequence of section \ref{sec_sequence} cropped to a length of $L=1000$ TRs on a
MAGNETOM Skyra 3T scanners (Siemens Healthcare, Erlangen, Germany). 
\captionsetup[subfigure]{labelformat=empty}
%put is back to default:
%\captionsetup[subfigure]{labelformat=parens}

As a reference, we also showed in the supporting document the results of the original MRF and AIR-MRF at a longer sequence length L=3000 TRs. 
Note that it is a reference but not the ground truth. Indeed, as can be observed with Brainweb digital phantom data (in section \ref{sec:csmrf_exp_brainweb}), results at L=3000 TRs are better than those from a shorter sequence length, but there still exists a significant error at L=3000 TRs. Note also that the water inside the NIST phantom results in flow artefacts, which explains the non-uniform signal in the background which just contains water.
%Note that it is a reference but should not be interpreted as a ground truth. In particular as can be observed in the Brainweb digital phantom experiment of section \ref{sec:csmrf_exp_brainweb}, results at L=3000 TRs are better than for a shorter sequence length but these improvements are relatively small compared to the error still existing at L=3000 TRs.
We can see in Figure \ref{fig_NIST_nonoise} that in the noise-free case, the other methods were able to correct the artefacts of original MRF method on the T2 map, and that our proposed \methodGFB\ method was able to further remove ringing artefacts. 
Note that  kt-SVD-MRF overestimated the PD map.

In the noisy case (Figure \ref{fig_NIST_with_noise}), our proposed \methodGFB\ method is the only method that was able to get rid of the noise and estimate maps that look similar to the reference.  
%Zoom on these images (both in noisy and noiseless case) are depicted in the supporting document.

%Some numerical analysis have also been performed on some tubes of the NIST phantom. In particular, 
The values and standard deviations (std) of the three regions of interest (ROI) corresponding to the three ROIs of the zoomed NIST image depicted in the supported document were calculated.
The most striking result was that the normalized std (i.e. std divided by the mean value) is around 4.5\% for T1 (resp. 8.5\% for T2) for the other methods and 3.4\% fot T1 (6.9\% for T2) for \methodGFB\ in the no-noise case. In the noisy case,
the normalized std was way higher for other methods: around 8.5\% for T1, (resp. between 22.4\% and 25.7\% for T2), while it staid at the same value of 3.5\% for T1 (6.3\% for T2) with \methodGFB.

    % mean(dataStdArr([3,4,13],:,1)./dataMeanArr([3,4,13],:,1)*100)
    % T1 with noise: 8.2206    8.4668    8.8970    3.5043
    % T2 with noise: 22.4100   23.3284   25.6575    6.3106
    % T1 no noise: 	4.3765    4.3619    4.5307    3.3535
    % T2 no noise: 8.3185    7.2594    8.4538    6.9501

%\begin{figure}[htb]
%\centering
%    \includegraphics[width=\linewidth]{figures/NIST/roi/T1Nist_nonoiseT1.png}
%  \caption{T1 values from T1 layer of the NIST phantom. Standard deviation is computed on the pixels inside each region of interest.}
%\end{figure}
%
%\begin{figure}[htb]
%\centering
%    \includegraphics[width=\linewidth]{figures/NIST/roi/T2Nist_nonoiseT2.png}
% \caption{T2 values from T2 layer of the NIST phantom. Standard deviation is computed on the pixels inside each region of interest.}
%\end{figure}

\subsection{Volunteer data evaluations\label{sec_volunteer}}

%%noise free case
%{
%\begin{table}[h]
%\begin{tabular}{cc}
%Piano player roll & Compact disc \\
%\includegraphics[height=2.5in]{figures/Volunteer/nonoise/result_SVDMRF_volunteerb1v22_L1000_k200/mrfT1.png} & 
%\includegraphics[height=2.5in]{figures/Volunteer/nonoise/result_SVDMRF_volunteerb1v22_L1000_k200/mrfT1.png} \\
%\end{tabular}
%\end{table}
%}
In order to test the different methods on a realistic data, a volunteer data with B1 map was acquired using the sequence described in section \ref{sec_sequence} (i.e. a sequence of  $L=3000$ TRs retrospectively cropped to a length of $L=1000$ TRs) on a MAGNETOM Prisma 3T MR scanner (Siemens Healthcare, Erlangen, Germany). % with the following parameters: FOV 300 $\times$ 300 mm$^2$, resolution 1.17  $\times$ 1.17  mm$^2$, slice thickness 5 mm. 
A 20-channel head coil was used. Coil sensitivity maps were calculated following the method in \cite{uecker2014espirit} using the k-space central region (12 $\times$ 12) after gridding the temporal averaged MRF data.

The only thing that changes when using the B1 map compared to the case where there is no B1 map, was in the dictionary matching step. Each B1 value has its own fingerprint dictionary, and for each voxel the dictionary corresponding to the B1 value of that voxel is choosen in order to perform the matching.

\begin{figure}[htb]
\centering
  \rotatebox[origin=l]{90}{\makebox[1.3in]{classical MRF}}%
  \hspace{0pt}
  \begin{subfigure}[b]{.32\linewidth}
    \centering
    \includegraphics[width=.99\textwidth]{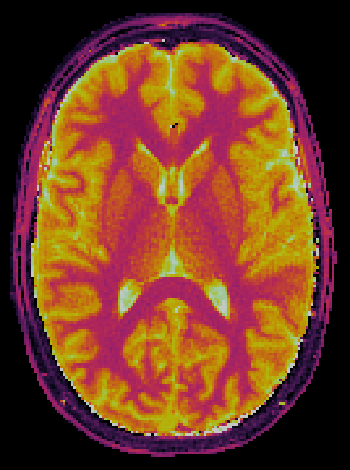}
  \end{subfigure}%   
  \begin{subfigure}[b]{.32\linewidth}
    \centering
    \includegraphics[width=.99\textwidth]{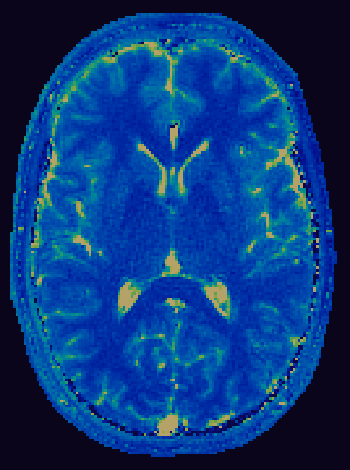}
  \end{subfigure}%  
    \begin{subfigure}[b]{.32\linewidth}
    \centering
    \includegraphics[width=.99\textwidth]{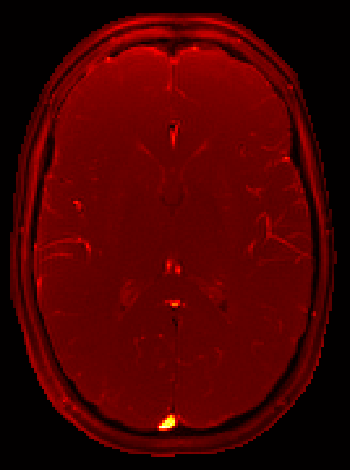}
  \end{subfigure}\\%
    \rotatebox[origin=l]{90}{\makebox[1.3in]{AIR-MRF}}%
  \hspace{0pt}
    \begin{subfigure}[b]{.32\linewidth}
    \centering
    \includegraphics[width=.99\textwidth]{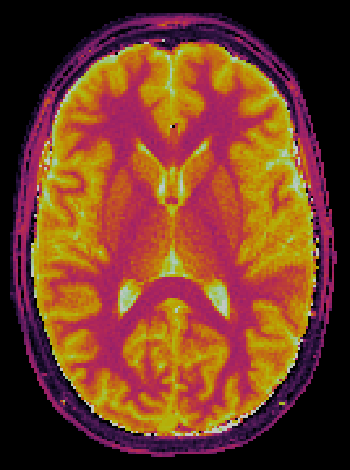}
  \end{subfigure}%   
  \begin{subfigure}[b]{.32\linewidth}
    \centering
    \includegraphics[width=.99\textwidth]{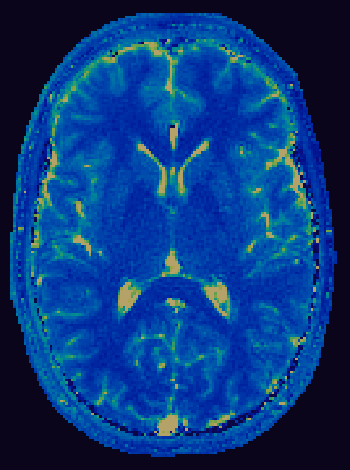}
  \end{subfigure}%  
    \begin{subfigure}[b]{.32\linewidth}
    \centering
    \includegraphics[width=.99\textwidth]{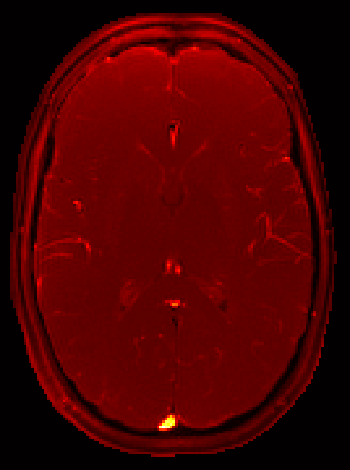}
  \end{subfigure}\\%
    \rotatebox[origin=l]{90}{\makebox[1.3in]{kt-SVD-MRF}}%
  \hspace{0pt}
   \begin{subfigure}[b]{.32\linewidth}
    \centering
    \includegraphics[width=.99\textwidth]{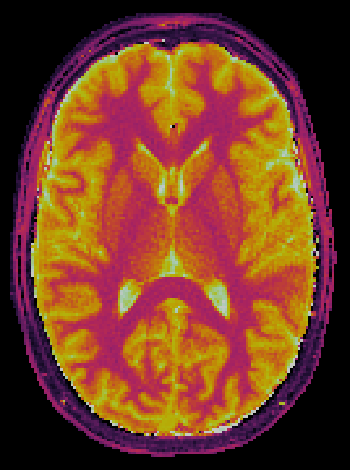}
  \end{subfigure}%   
  \begin{subfigure}[b]{.32\linewidth}
    \centering
    \includegraphics[width=.99\textwidth]{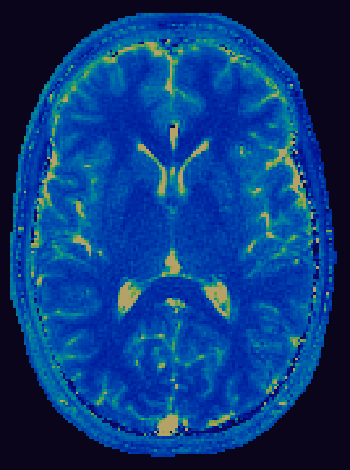}
  \end{subfigure}%  
    \begin{subfigure}[b]{.32\linewidth}
    \centering
    \includegraphics[width=.99\textwidth]{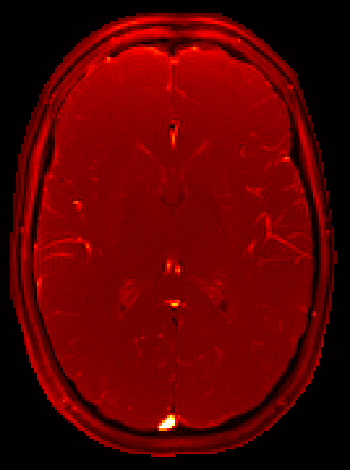}
  \end{subfigure}\\%
    \rotatebox[origin=l]{90}{\makebox[1.3in]{\methodGFB}}%
  \hspace{0pt}
      \begin{subfigure}[b]{.32\linewidth}
    \centering
    \includegraphics[width=.99\textwidth]{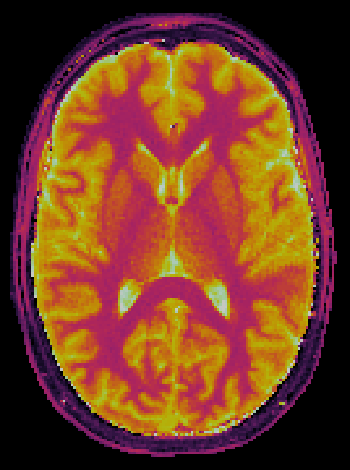}
  \end{subfigure}%   
  \begin{subfigure}[b]{.32\linewidth}
    \centering
    \includegraphics[width=.99\textwidth]{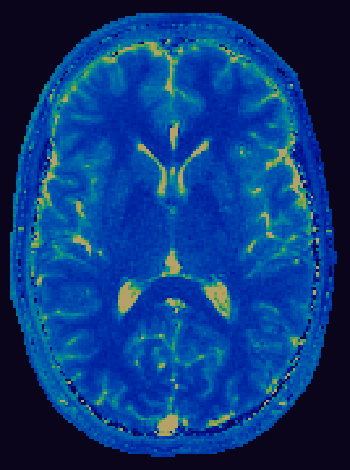}
  \end{subfigure}%  
    \begin{subfigure}[b]{.32\linewidth}
    \centering
    \includegraphics[width=.99\textwidth]{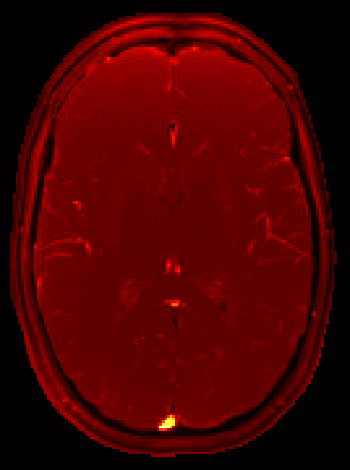}
  \end{subfigure}\\%
    \rotatebox[origin=l]{90}{\makebox[.5in]{\quad}}%
  \hspace{0pt}
  \begin{subfigure}[b]{.32\linewidth}
    \centering
    \includegraphics[width=.99\textwidth]{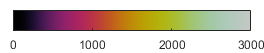}
    \caption{T1}\label{fig:1a}
  \end{subfigure}%   
  \begin{subfigure}[b]{.32\linewidth}
    \centering
    \includegraphics[width=.99\textwidth]{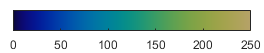}
    \caption{T2}\label{fig:1b}
  \end{subfigure}%  
    \begin{subfigure}[b]{.32\linewidth}
    \centering
    \includegraphics[width=.99\textwidth]{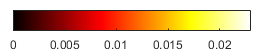}
    \caption{PD}\label{fig:1b}
  \end{subfigure}%
  \caption{Volunteer data evaluations with noise-free measurements and sequence length 1000}\label{fig_volunteer_nonoise}
\end{figure}

\begin{figure}[htb]
\centering
  \rotatebox[origin=l]{90}{\makebox[1.3in]{classical MRF}}%
  \hspace{0pt}
  \begin{subfigure}[b]{.32\linewidth}
    \centering
    \includegraphics[width=.99\textwidth]{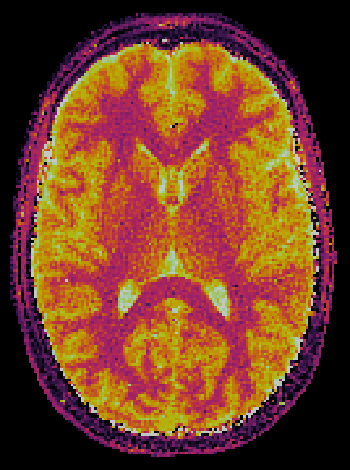}
  \end{subfigure}%   
  \begin{subfigure}[b]{.32\linewidth}
    \centering
    \includegraphics[width=.99\textwidth]{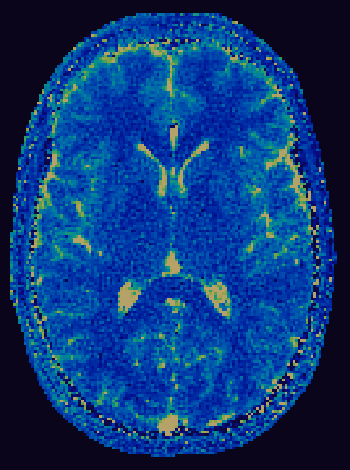}
  \end{subfigure}%  
    \begin{subfigure}[b]{.32\linewidth}
    \centering
    \includegraphics[width=.99\textwidth]{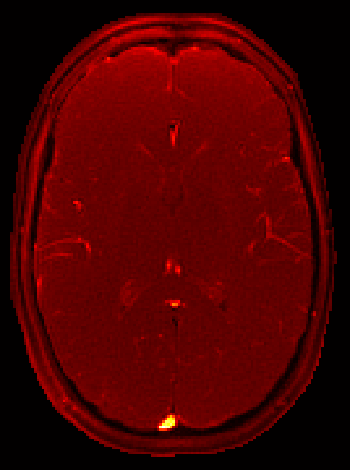}
  \end{subfigure}\\%
    \rotatebox[origin=l]{90}{\makebox[1.3in]{AIR-MRF}}%
  \hspace{0pt}
    \begin{subfigure}[b]{.32\linewidth}
    \centering
    \includegraphics[width=.99\textwidth]{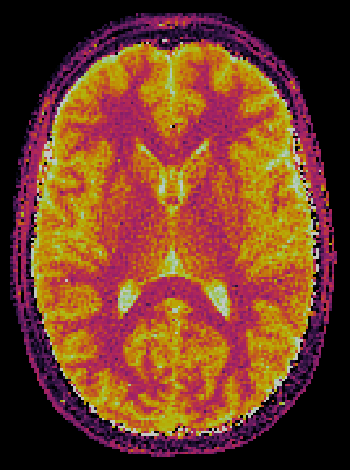}
  \end{subfigure}%   
  \begin{subfigure}[b]{.32\linewidth}
    \centering
    \includegraphics[width=.99\textwidth]{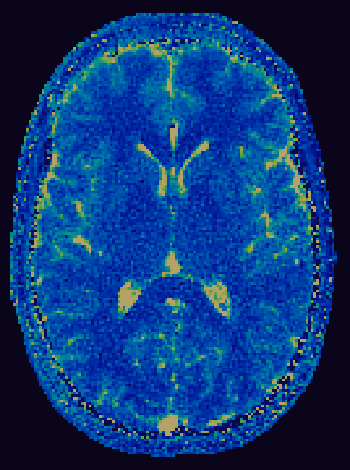}
  \end{subfigure}%  
    \begin{subfigure}[b]{.32\linewidth}
    \centering
    \includegraphics[width=.99\textwidth]{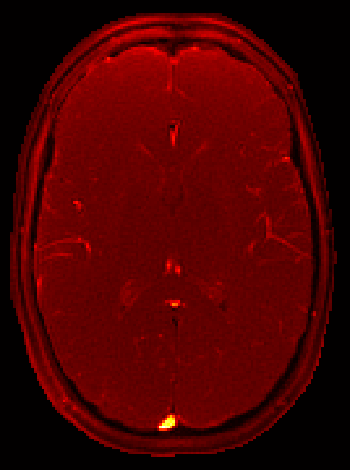}
  \end{subfigure}\\%
    \rotatebox[origin=l]{90}{\makebox[1.3in]{kt-SVD-MRF}}%
  \hspace{0pt}
   \begin{subfigure}[b]{.32\linewidth}
    \centering
    \includegraphics[width=.99\textwidth]{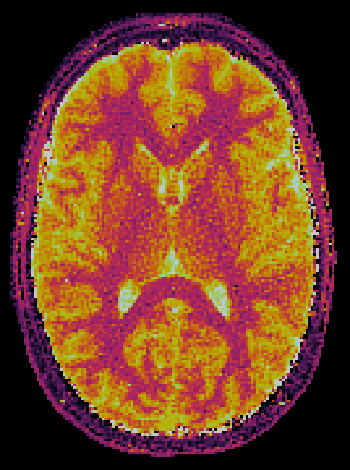}
  \end{subfigure}%   
  \begin{subfigure}[b]{.32\linewidth}
    \centering
    \includegraphics[width=.99\textwidth]{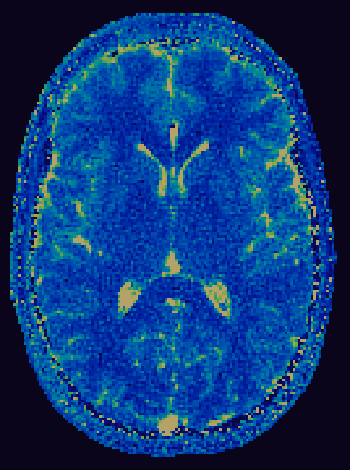}
  \end{subfigure}%  
    \begin{subfigure}[b]{.32\linewidth}
    \centering
    \includegraphics[width=.99\textwidth]{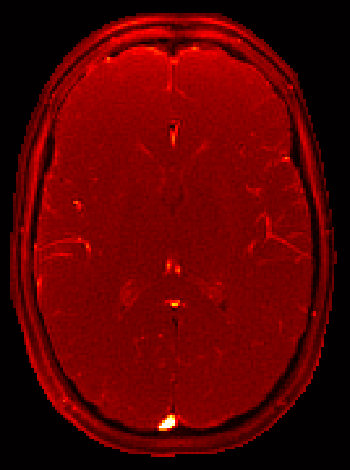}
  \end{subfigure}\\%
    \rotatebox[origin=l]{90}{\makebox[1.3in]{\methodGFB}}%
  \hspace{0pt}
      \begin{subfigure}[b]{.32\linewidth}
    \centering
    \includegraphics[width=.99\textwidth]{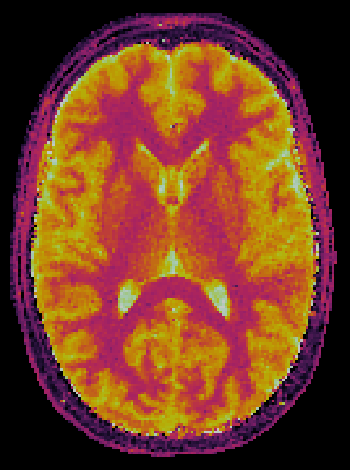}
  \end{subfigure}%   
  \begin{subfigure}[b]{.32\linewidth}
    \centering
    \includegraphics[width=.99\textwidth]{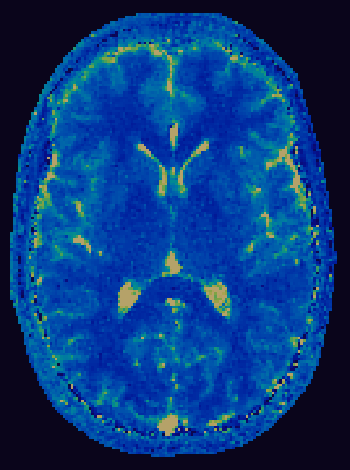}
  \end{subfigure}%  
    \begin{subfigure}[b]{.32\linewidth}
    \centering
    \includegraphics[width=.99\textwidth]{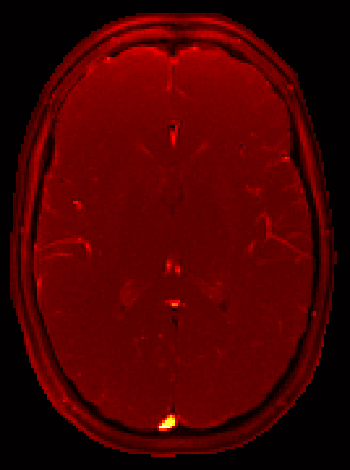}
  \end{subfigure}\\%
            \rotatebox[origin=l]{90}{\makebox[.5in]{\quad}}%
  \hspace{0pt}
  \begin{subfigure}[b]{.32\linewidth}
    \centering
    \includegraphics[width=.99\textwidth]{figures/colormapT1Volunteer.png}
    \caption{T1}\label{fig:1a}
  \end{subfigure}%   
  \begin{subfigure}[b]{.32\linewidth}
    \centering
    \includegraphics[width=.99\textwidth]{figures/colormapT2Volunteer.png}
    \caption{T2}\label{fig:1b}
  \end{subfigure}%  
    \begin{subfigure}[b]{.32\linewidth}
    \centering
    \includegraphics[width=.99\textwidth]{figures/colormapPDVolunteer.png}
    \caption{PD}\label{fig:1b}
  \end{subfigure}%
  \caption{Volunteer data evaluations with noisy measurements and sequence length 1000}\label{fig:volunteer_noise}
\end{figure}

Results  on volunteer data without added noise are depicted in Figure \ref{fig_volunteer_nonoise}.
Results of the other methods look similar to each other, but results of our proposed \methodGFB\ method look slightly smoother.
Results on noisy data (cf section \ref{sec:noisycase}), depicted in Figure \ref{fig:volunteer_noise}, show that \methodGFB\ was able to remove the noise and obtained smoother maps, more similar to the noise-free reference maps than the other methods.

\section{Discussion}

We presented a new method for MRF which exploits jointly low-rank (via SVD comprression), spatial (via TV regularization) and temporal regularizations (via Bloch manifold projection, i.e. fingerprint matching), into an iterative algorithm, to improve MRF reconstruction.

We presented and evaluated different variants of our algorithm.
Experiments showed that better results were obtained with back-tracking step size compared to fixed step size (cf comparison between \methodXXIV\ and \methodXXIV  BT),
that  better results were obtained with matching at every iteration as opposed to not macthing at every iteration (cf comparison between  \methodXXIV  BT and \methodXXXIV).
Experiments did not indicate that the low-rank approach via autocalibration was better than the classical approach consisting of just using the dictionary for the SVD compression.
In a first variant the spatial regularization and the fingerprint matching are applied sequencially (IGP-MRF), in the second variant they are applied in parallel (\methodGFB).
Experiments showed that \methodGFB\ obtained better results than IGP-MRF. 

In the second part of the experiments, we only compared our best method i.e. \methodGFB\ 
w.r.t.  some methods of the state of the art.
Experiments on real data, showed that all the tested methods performed relatively well at a short sequence length of 1000 TRs, but only our proposed method was able to get rid of the noise when the measurements were contaminated with noise.

In terms of computational performance, all the methods were implemented in Matlab and were executed on a windows 7 platform with a dual CPU Intel Xeon 2.3GHz and 64 GB of RAM.
Our proposed method took, for the Brainweb dataset, about 20s per iteration, plus an additional 8.5s overhead for the initialization, so for 10 iterations the overall runtime was about 210s in Matlab. It is slightly worse than kt-SVD-MRF which took 170s with our Matlab implementation.
However, in terms of memory, our method took around only 5MB of memory to store the MRF data as opposed to around 768MB for kt-SVD-MRF because in our method, the data size is a matrix of size: number of pixels x dictionary compression ratio $k$, (i.e. roughly $(256)^2 \times 10 = 655'360$) which doesn't depend on the sequence length, while the data size of kt-SVD-MRF is: sequence length $L$ x readout length x number of spirals (i.e. roughly $1000 \times 2000 \times 48 = 96'000'000$).

% discussion
% 1) main results
% 2) limitations
% 3) Future work
% 4) autocalibration: not so useful
% 5) 
% step sizes: BT better than FSZ in noisy case.  IGP-MRF-01 worst than IGP-MRF-01BT in the noisy case
% autocalibration: not useful in practice in our experiments (comparisons of  IGP-MRF-10 vs IGP-MRF-00).
% matching at every iteration: useful (comparisons of  IGP-MRF-01BT vs IGP-MRF-00)
% best: GFB, -> explain why we only compare GFB to the soa
% minimum sequence length ?
% computational efficiency

\section{Conclusion}

We presented a new method for MRF which is able to exploit efficiently low-rank, spatial and temporal regularizations, into an iterative algorithm, to improve MRF reconstruction.
In order to regularize the images both spatially and temporally (via the Bloch manifold constraint), our algorithm applies these two regularizations in parallel as suggested by the generalized forward-backward algorithm.
On the other hand, the low-rank regularization is imposed via an offline SVD compression so that no SVD is performed during the execution of the algorithm.
The superiority of the proposed approach compared to state-of-the-art methods was demonstrated on Brainweb digital phantom data and scanned data.
On Brainweb digital phantom data, we show that the average error of our proposed method is about two times smaller than the other methods for all the tested cases, i.e. for different sequence lengths (from 600 to 3000 TRs), with or without additional noise to the measurements.
While the improvement on low-noise scanned data is difficult to assess, the proposed method was the only tested method that was able to reconstruct good-quality maps in case of measurements corrupted by noise.
%The difference on real noise-free data was however more teneous. 

We surmise that the gap in the performance between the results on digital phantom data and scanned data is due to the forward imaging model which do not take into account approximations in the k-space trajectory, motion, chemical shift, magnetic susceptibility, inhomogeneous B0 and B1 fields, etc.
As a consequence, we think that the proposed method could be further improved by improving the forward imaging model used in our algorithm.
One way to improve the model would be to replace some of the operators with trainable modules %(e.g. using a leanable FFT \cite{moreau2018}) 
that learn to correct the model imperfections.
%As an example, the FFT could be replaced by a leanable FFT \cite{moreau2018}.
%However, the main challenge with deep learning approaches for MRF is that the training (particularly the ground truth) data is not easily available and using synthetic data for training will not solve the problem of model approximation as mentioned earlier.

\bibliography{article}

\begin{thebibliography}{10}
\expandafter\ifx\csname url\endcsname\relax
  \def\url#1{\texttt{#1}}\fi
\expandafter\ifx\csname urlprefix\endcsname\relax\def\urlprefix{URL }\fi
\expandafter\ifx\csname href\endcsname\relax
  \def\href#1#2{#2} \def\path#1{#1}\fi

\bibitem{ma2013magnetic}
D.~Ma, V.~Gulani, N.~Seiberlich, K.~Liu, J.~L. Sunshine, J.~L. Duerk, M.~A.
  Griswold, Magnetic resonance fingerprinting, Nature 495~(7440) (2013)
  187--192.

\bibitem{zhao2016optimal}
B.~Zhao, J.~P. Haldar, K.~Setsompop, L.~L. Wald, Optimal experiment design for
  magnetic resonance fingerprinting, in: 2016 38th Annual International
  Conference of the IEEE Engineering in Medicine and Biology Society (EMBC),
  IEEE, 2016, pp. 453--456.

\bibitem{mcgivney2014svd}
D.~F. McGivney, E.~Pierre, D.~Ma, Y.~Jiang, H.~Saybasili, V.~Gulani, M.~A.
  Griswold, {SVD} compression for magnetic resonance fingerprinting in the time
  domain, IEEE transactions on medical imaging 33~(12) (2014) 2311--2322.

\bibitem{cao2017robust}
X.~Cao, C.~Liao, Z.~Wang, Y.~Chen, H.~Ye, H.~He, J.~Zhong, Robust
  sliding-window reconstruction for accelerating the acquisition of mr
  fingerprinting, Magnetic resonance in medicine 78~(4) (2017) 1579--1588.

\bibitem{davies2014compressed}
M.~Davies, G.~Puy, P.~Vandergheynst, Y.~Wiaux, A compressed sensing framework
  for magnetic resonance fingerprinting, SIAM Journal on Imaging Sciences 7~(4)
  (2014) 2623--2656.

\bibitem{cline2017air}
C.~C. Cline, X.~Chen, B.~Mailh{\'e}, Q.~Wang, J.~Pfeuffer, M.~Nittka, M.~A.
  Griswold, P.~Speier, M.~S. Nadar, {AIR-MRF}: Accelerated iterative
  reconstruction for magnetic resonance fingerprinting, Magnetic resonance
  imaging 41 (2017) 29--40.

\bibitem{wang2016magnetic}
Z.~Wang, H.~Li, Q.~Zhang, J.~Yuan, X.~Wang, Magnetic resonance fingerprinting
  with compressed sensing and distance metric learning, Neurocomputing 174
  (2016) 560--570.

\bibitem{zhao2016maximum}
B.~Zhao, K.~Setsompop, H.~Ye, S.~F. Cauley, L.~L. Wald, Maximum likelihood
  reconstruction for magnetic resonance fingerprinting, IEEE transactions on
  medical imaging 35~(8) (2016) 1812--1823.

\bibitem{pierre2016multiscale}
E.~Y. Pierre, D.~Ma, Y.~Chen, C.~Badve, M.~A. Griswold, Multiscale
  reconstruction for mr fingerprinting, Magnetic resonance in medicine 75~(6)
  (2016) 2481--2492.

\bibitem{doneva2017matrix}
M.~Doneva, T.~Amthor, P.~Koken, K.~Sommer, P.~B{\"o}rnert, Matrix
  completion-based reconstruction for undersampled magnetic resonance
  fingerprinting data, Magnetic Resonance Imaging 41 (2017) 41--52.

\bibitem{zhao2017improved}
B.~Zhao, K.~Setsompop, E.~Adalsteinsson, B.~Gagoski, H.~Ye, D.~Ma, Y.~Jiang,
  P.~Ellen~Grant, M.~A. Griswold, L.~L. Wald, Improved magnetic resonance
  fingerprinting reconstruction with low-rank and subspace modeling, Magnetic
  Resonance in Medicine 79~(2) (2018) 933--942.

\bibitem{mazor2016low}
G.~Mazor, L.~Weizman, A.~Tal, Y.~C. Eldar, Low rank magnetic resonance
  fingerprinting, in: 2016 38th Annual International Conference of the IEEE
  Engineering in Medicine and Biology Society (EMBC), IEEE, 2016, pp. 439--442.

\bibitem{asslander2018low}
J.~Assl{\"a}nder, M.~A. Cloos, F.~Knoll, D.~K. Sodickson, J.~Hennig,
  R.~Lattanzi, Low rank alternating direction method of multipliers
  reconstruction for mr fingerprinting, Magnetic resonance in medicine 79~(1)
  (2018) 83--96.

\bibitem{bustin2019high}
A.~Bustin, G.~Lima~da Cruz, O.~Jaubert, K.~Lopez, R.~M. Botnar, C.~Prieto,
  High-dimensionality undersampled patch-based reconstruction (hd-prost) for
  accelerated multi-contrast mri, Magnetic resonance in medicine 81~(6) (2019)
  3705--3719.

\bibitem{lima2019sparsity}
G.~Lima~da Cruz, A.~Bustin, O.~Jaubert, T.~Schneider, R.~M. Botnar, C.~Prieto,
  Sparsity and locally low rank regularization for mr fingerprinting, Magnetic
  resonance in medicine 81~(6) (2019) 3530--3543.

\bibitem{arberet2017}
S.~Arberet, X.~Chen, B.~Mailh{\'e}, P.~Speier, M.~S. Nadar, {CS-MRF}: Sparse \&
  low-rank iterative reconstruction for magnetic resonance fingerprinting, in:
  ISMRM Workshop on Magnetic Resonance Fingerprinting, 2017.

\bibitem{arberetGFB2019}
S.~Arberet, X.~Chen, B.~Mailh{\'e}, P.~Speier, M.~Nittka, H.~Meyer, M.~S.
  Nadar, {GFB-MRF}: Parallel spatial and bloch manifold regularized iterative
  reconstruction for magnetic resonance fingerprinting, in: Proc. 27th Sci.
  Meet. Int. Soc. Magn. Reson. Med., 2019.

\bibitem{bertsekas2011incremental}
D.~P. Bertsekas, Incremental gradient, subgradient, and proximal methods for
  convex optimization: A survey, Optimization for Machine Learning 2010~(1-38)
  (2011) 3.

\bibitem{raguet2013generalized}
H.~Raguet, J.~Fadili, G.~Peyr{\'e}, A generalized forward-backward splitting,
  SIAM Journal on Imaging Sciences 6~(3) (2013) 1199--1226.

\bibitem{chambolle2004algorithm}
A.~Chambolle, An algorithm for total variation minimization and applications,
  Journal of Mathematical imaging and vision 20~(1) (2004) 89--97.

\bibitem{kamilov2014variational}
U.~S. Kamilov, E.~Bostan, M.~Unser, Variational justification of cycle spinning
  for wavelet-based solutions of inverse problems, IEEE Signal Processing
  Letters 21~(11) (2014) 1326--1330.

\bibitem{mailhe2018fast}
B.~Mailh{\'e}, A.~Ruppel, Q.~Wang, M.~S. Nadar, Fast and memory efficient
  redundant wavelet regularization with sequential cycle spinning, {US} Patent
  9,858,689 (Jan.~2 2018).

\bibitem{cauley2015fast}
S.~F. Cauley, K.~Setsompop, D.~Ma, Y.~Jiang, H.~Ye, E.~Adalsteinsson, M.~A.
  Griswold, L.~L. Wald, Fast group matching for {MR} fingerprinting
  reconstruction, Magnetic resonance in medicine 74~(2) (2015) 523--528.

\bibitem{muja2009flann}
M.~Muja, D.~Lowe, Flann-fast library for approximate nearest neighbors user
  manual, Computer Science Department, University of British Columbia,
  Vancouver, BC, Canada (2009).

\bibitem{cohen2017deep}
O.~Cohen, B.~Zhu, M.~Rosen, Deep learning for fast mr fingerprinting
  reconstruction, in: 2017 Scientific Meeting Proceedings. International
  Society for Magnetic Resonance in Medicine, 2017, p. 688.

\bibitem{fang2019deep}
Z.~Fang, Y.~Chen, M.~Liu, L.~Xiang, Q.~Zhang, Q.~Wang, W.~Lin, D.~Shen, Deep
  learning for fast and spatially constrained tissue quantification from highly
  accelerated data in magnetic resonance fingerprinting, IEEE transactions on
  medical imaging 38~(10) (2019) 2364--2374.

\bibitem{song2019magnetic}
P.~Song, Y.~C. Eldar, G.~Mazor, M.~R. Rodrigues, Magnetic resonance
  fingerprinting using a residual convolutional neural network, in: ICASSP
  2019-2019 IEEE International Conference on Acoustics, Speech and Signal
  Processing (ICASSP), IEEE, 2019, pp. 1040--1044.

\bibitem{nistphantom}
{High Precision Devices Inc.},
  \href{http://www.hpd-online.com/MRI-phantoms.php}{{NIST} phantom}, {B}oulder,
  Colorado, USA (2016).
\newline\urlprefix\url{http://www.hpd-online.com/MRI-phantoms.php}

\bibitem{fessler2003nonuniform}
J.~A. Fessler, B.~P. Sutton, Nonuniform fast fourier transforms using min-max
  interpolation, IEEE transactions on signal processing 51~(2) (2003) 560--574.

\bibitem{recht2010guaranteed}
B.~Recht, M.~Fazel, P.~A. Parrilo, Guaranteed minimum-rank solutions of linear
  matrix equations via nuclear norm minimization, SIAM review 52~(3) (2010)
  471--501.

\bibitem{combettes2011proximal}
P.~L. Combettes, J.-C. Pesquet, Proximal splitting methods in signal
  processing, in: Fixed-point algorithms for inverse problems in science and
  engineering, Springer, 2011, pp. 185--212.

\bibitem{blumensath2010normalized}
T.~Blumensath, M.~E. Davies, Normalized iterative hard thresholding: Guaranteed
  stability and performance, IEEE Journal of selected topics in signal
  processing 4~(2) (2010) 298--309.

\bibitem{cocosco1997brainweb}
C.~A. Cocosco, V.~Kollokian, R.~K.-S. Kwan, G.~B. Pike, A.~C. Evans, Brainweb:
  Online interface to a 3d {MRI} simulated brain database, NeuroImage 5~(4)
  (1997) 301--307.

\bibitem{ktsvdmrf2017}
E.~Pierre, M.~A. Griswold, A.~Connelly, Fast analytical solution for extreme
  unaliasing of {MRF} image series, in: Proc. 25th Sci. Meet. Int. Soc. Magn.
  Reson. Med., 2017.

\bibitem{meyer2011dual}
C.~Meyer, L.~Zhao, M.~Lustig, M.~Jilwan-Nicolas, M.~Wintermark, J.~Mugler,
  F.~Epstein, Dual-density and parallel spiral asl for motion artifact
  reduction, in: Proc. Intl. Soc. Mag. Reson. Med, Vol.~19, 2011, p. 3986.

\bibitem{pfeuffer2017mitigation}
J.~Pfeuffer, A.~Kechagias, G.~K{\"o}rzd{\"o}rfer, D.~Ma, M.~Griswold,
  M.~Nittka, Mitigation of spiral undersampling artifacts in magnetic resonance
  fingerprinting (mrf) by adapted interleaf reordering, in: Proceedings of the
  25th Annual Meeting of ISMRM, Honolulu, Vol. 133, 2017.

\bibitem{chen2001atomic}
S.~S. Chen, D.~L. Donoho, M.~A. Saunders, Atomic decomposition by basis
  pursuit, SIAM review 43~(1) (2001) 129--159.

\bibitem{griswoldMRFcolor2018}
M.~Griswold, J.~Sunshine, N.~Seiberlich, V.~Gulani, Towards unified colormaps
  for quantitative {MRF} data, in: Proc. 26th Sci. Meet. Int. Soc. Magn. Reson.
  Med., 2018.

\bibitem{uecker2014espirit}
M.~Uecker, P.~Lai, M.~J. Murphy, P.~Virtue, M.~Elad, J.~M. Pauly, S.~S.
  Vasanawala, M.~Lustig, {ESPIRiT-An} eigenvalue approach to autocalibrating
  parallel {MRI}: where {SENSE} meets {GRAPPA}, Magnetic resonance in medicine
  71~(3) (2014) 990--1001.

\end{thebibliography}

\end{document}